\documentclass[man, 11, floatsintext]{apa7}

\usepackage[american]{babel}
\usepackage[T1]{fontenc} 
\usepackage{mathptmx}
\usepackage{url}
\usepackage{amsmath}
\usepackage{amssymb}
\usepackage{caption}
\usepackage{graphicx}
\usepackage{tikz}
\usepackage{orcidlink}
\usepackage[frozencache=true,cachedir=minted-cache]{minted} 
\usepackage{enumitem}
\usepackage{makecell}
\usepackage{adjustbox}
\usepackage{multirow}
\usepackage{mathtools}
\usepackage{csquotes}
\usepackage{xfrac}
\usepackage{bm}
\usepackage{setspace}
\usepackage[style=apa, citestyle=authoryear-comp, uniquelist=false, sorting=nyt, backend=biber, uniquename=false, hyperref=true, sortcites=false]{biblatex}
\usepackage{appendix}

\DeclareLanguageMapping{american}{american-apa}
\DeclareUnicodeCharacter{0302}{\^{R}}
\DeclareDelimFormat{nameyeardelim}{\addcomma\space}
\DeclareMathAlphabet{\pazocal}{OMS}{zplm}{m}{n}
\SetMathAlphabet\pazocal{bold}{OMS}{zplm}{bx}{n}

\title{Bayesian Multilevel Compositional Data Analysis: \\Introduction, Evaluation, and Application}
\shorttitle{Bayesian Multilevel Compositional Data Analysis}
\author{
Flora Le\orcidlink{0000-0003-0089-8167}$^1$, 
Tyman E. Stanford\orcidlink{0000-0002-8570-5493}$^2$, 
Dorothea Dumuid\orcidlink{0000-0003-3057-0963}$^2$, and 
Joshua F. Wiley\orcidlink{0000-0002-0271-6702}$^1$ 
}

\authorsaffiliations{
$^1$ School of Psychological Sciences{,} ,
Monash University
\\
$^2$ Alliance for Research in Exercise{,} Nutrition and Activity{,},
Allied Health and Human Performance{,} ,
University of South Australia
}

\authornote{
All analysis code for this study is available at: 
\url{https://github.com/florale/multilevelcoda-sim}.
Materials for the real data study are available on the Open Science Framework (\url{https://doi.org/10.17605/OSF.IO/H5497}\nocite{aces}, 
\url{https://doi.org/10.17605/OSF.IO/QM63W}\nocite{destress}, and 
\url{https://doi.org/10.17605/OSF.IO/TZ48Y}\nocite{shs}).
Data for the real data study are available from the corresponding authors upon request.
\\
\textbf{Author Contributions}.
FL: Conceptualization, data curation, formal analysis, investigation, methodology, project administration, software, visualization, writing-original draft, writing-review and editing.
TES: Conceptualization, resources, supervision, writing-review and editing.
DD: Conceptualization, resources, supervision, writing-review and editing.
JFW: Conceptualization, data curation, investigation, methodology, project administration, resources, software, supervision, writing-review and editing.
\\
\textbf{Acknowledgements}. 
FL was supported by Monash University through a Monash Graduate Scholarship and a Monash International Tuition Scholarship. DD was funded by an Australian Research Council (ARC) Discovery Early Career Award (DECRA) DE230101174 and by the Centre of Research Excellence in Driving Global Investment in Adolescent Health funded by National Health and Medical Research Council (NHMRC) GNT1171981. JFW was supported by a NHMRC fellowship (1178487). The authors declared no conflict of interests.
\\
Correspondence to 
Flora Le (\url{flora.le@monash.edu}) or 
Joshua F. Wiley (\url{joshua.wiley@monash.edu}),
School of Psychological Sciences, 
Monash University, Clayton, VIC, Australia.
}
\leftheader{Le, Stanford, Dumuid, and Wiley}

\abstract{
Multilevel compositional data are data that are repeatedly measured or clustered within groups and are non-negative and sum to a constant value. 
These data arise in various settings, such as intensive, longitudinal studies using ecological momentary assessments and wearable devices.
Examples include 24h sleep-wake behaviours, sleep architecture, and macronutrients.
This article presents a novel method for analysing multilevel compositional data using Bayesian inference.
We describe the theoretical details of the data and the models,
and outline the steps necessary to implement this method.
We introduce the \textbf{R} package \textit{multilevelcoda} to facilitate the application of this method and illustrate using a real data example.
An extensive parameter recovery simulation study verified the robust performance of the method.
Across all conditions investigated in the simulation study, 
the fitted models had minimal convergence issues (convergence rate $>$ 99\%) and
achieved excellent quality parameter estimates and inference, 
with an average bias of 0.00 (range -0.09, 0.05)  
and coverage of 0.95 (range 0.93, 0.97).
We conclude the article with recommendations on the use of the 
Bayesian multilevel compositional data analysis. We
hope to promote wider application of this method to
gain novel and robust answers to scientific questions.
}

\begin{document}
\maketitle % This tells LaTeX to make the title page
%\section{Introduction}
Multilevel data are increasingly collected in many fields, including psychology.
Common types of multilevel data such as
24-hour sleep-wake behaviours (e.g., time spent in sleep, physical activity, and 
sedentary behaviour, during the 24h day) and  
macronutrients (e.g., proportions of total caloric intake from 
macronutrients like proteins, fats and carbohydrates) have a 
compositional structure. Data are compositional when they consist of parts
that contain relative information about the whole, which are represented
as non-negative values that sum to a constant.
Compositional data can be expressed as percentages (or proportions) or
in other units that are constrained to a total constant value
(e.g., 1440 minutes in a day). 
The constrained, constant-sum nature of compositions imposes perfect multi-collinearity among the components, causing the covariance structure of the data to be negatively biased \parencite{aitchison1982}.
Accordingly, standard statistical methods, such as linear models, are not appropriate for fitting raw compositional data and produce invalid results.

Compositional data analysis \parencite[CoDA;][]{aitchison1982},
originally developed for the analysis of geochemical data, has been increasingly employed outside of psychology. 
For example, behavioural epidemiology has shifted from considering individual behaviours
(e.g., sleep, physical activity, and sedentary behaviour) to consider behaviours as an
integrated 24h composition. This paradigm shift is demonstrated by the 
increasing use of CoDA \parencite{dumuid2018, dumuid2019, dumuid2020} to investigate how the reallocations of time across behaviours are associated with health outcomes \parencite{janssen2020, grgic2018, miatke2023}. Similarly, public guidelines have shifted to provide recommendations on 24h behaviours, rather than separate guidelines for each behaviour.
The current evidence base is, however, mostly cross-sectional \parencite{miatke2023}.
Longitudinal evidence remains limited, due to challenges and a lack of tools to analyse 
multilevel compositional data. Emerging advanced statistical methods that accommodate
the theoretical properties of multilevel compositional data could, therefore, 
facilitate more robust and conceptually meaningful inference, 
leading to improved health insights. In psychology where ecological momentary assessments (EMAs) 
and wearables are central methods in intensive, longitudinal studies,
such statistical methods can advance our current knowledge base on how real-time phenomena,
such as health behaviours, cognition and emotion, interact in everyday life.

Bayesian multilevel models offer flexibility in modelling statistical
phenomena that exist in different levels.
Although both Bayesian and frequentist models can include population- and group-level effects 
(commonly referred to as \textit{fixed} and \textit{random} effects),
Bayesian multilevel models are increasingly employed due to their flexibility and 
increases in computational capacity.
Further, advances in software for Bayesian posterior sampling,
including the probabilistic programming language \textbf{Stan} \parencite{carpenter2017, stan2023} 
and the \textbf{R} package \textit{brms} \parencite{brms, burkner2017}
with a front-end that requires minimal programming with similar syntax to 
frequentist multilevel models, \parencite[i.e., \textit{lme4},][]{lme4} have increased the popularity in 
Bayesian multilevel models. Finally, Bayesian multilevel models enable computationally easy and robust calculation of significance and uncertainty
intervals around predictions and other post model estimation quantities, even when
non-linear transformations are applied. This feature is particularly helpful 
for multilevel CoDA by enhancing the ease of reporting and interpreting results.

In this article, we present a novel method for multilevel CoDA 
using Bayesian inference. We start by describing the structure
of multilevel compositional data and the modelling approach for this data type. 
We then discuss the use of Bayesian inference for multilevel models using compositional variables.
We provide the multilevel models specification, with a focus on models with compositional variables 
as predictors. Next we introduce the substitution analysis to examine the 
reallocations between compositional parts associated with an outcome for 
easy model interpretation. To facilitate the implementation of multilevel CoDA
in a robust and principled workflow, we introduce the \textbf{R} package \textit{multilevelcoda} \parencite{multilevelcoda, le2024b}.
We illustrate \textit{multilevelcoda} on a data set with daily repeated measures.
We then use the results from the real data application as a starting point 
for a Monte Carlo simulation study to assess the accuracy and coverage 
of parameter estimates. We conclude the paper with a discussion about 
Bayesian multilevel CoDA and recommendations on its practical applications.

\section{Modelling Multilevel Compositional Data}
\subsection{Examples of Multilevel Compositional Data}
In this section, we introduce examples of compositional data that often arise in psychology. These compositional data become
multilevel when the sampling units are repeatedly measured (longitudinal data) or they are 
clustered within groups (hierarchical data). 

\subsubsection{Sleep-wake Behaviours}
\begin{figure*}[!htbp]
\centering
  \caption{An example composition of time spent in 24h behaviours
  of an individual is shown in Panel A. 
  Due to the fixed 24-hour day, 
  an individual can reallocate time across behaviours differently, 
  but they must keep the total time fixed. 
  For example, they may increase an hour of
  moderate-to-vigorous physical activity at the expense of sleep (Panel B).
  Alternatively, they may increase an hour of
  moderate-to-vigorous physical activity at the expense of sedentary behaviour (Panel C).
  }
  \includegraphics[width=\textwidth]{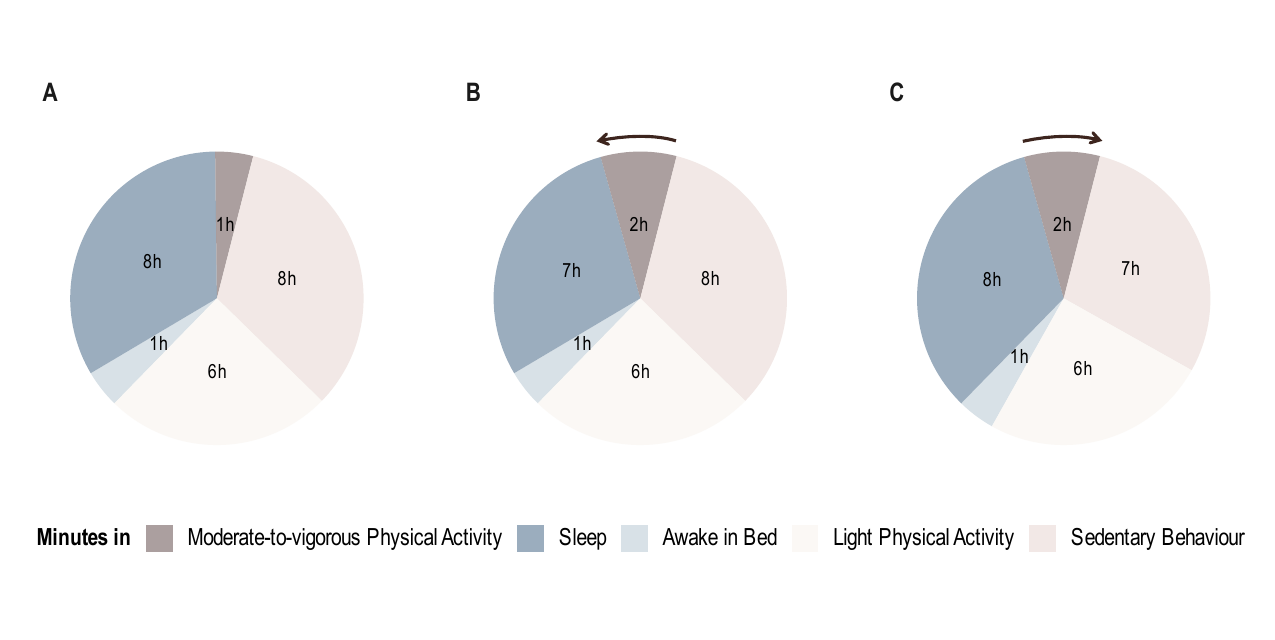}
\label{fig-comp}
\end{figure*}

\textbf{24h Behaviours.} Time spent in the 24h day can be categorised into multiple, 
mutually exclusive behaviours. From a lifestyle perspective, we 
can categorise behaviours into, for example: total sleep time, 
awake in bed (sleep onset latency and wake time after sleep onset), 
moderate-to-vigorous physical activity (MVPA), light physical activity (LPA), 
and sedentary behaviour (SB) \parencite{le2022}. Due to the fixed 24 hours in a day, 
a person cannot increase time spent in one behaviour while keeping all other 
behaviours and the total time fixed. An increased time spent in one behaviour 
must be compensated by an equal time decrease in one or more of the other behaviours. 
For example, a person can only increase time spent in physical activity 
by spending less time in other behaviours (e.g., sleep, sedentary behaviour), as illustrated in Figure \ref{fig-comp}.
Therefore, time spent in the 24h day is compositional data; 
the relative time spent in different behaviours is informative.
Behavioural time-use data are also often multilevel, due to the rise in 
passive wearable sensors making it easy to measure activity and sleep 
repeatedly over consecutive days. Although growing evidence exists on the 
associations between 24h behaviour composition and health outcomes \parencite{janssen2020, grgic2018, miatke2023}, 
analyses have mostly averaged the data across days to examine the 
cross-sectional associations. Insights from longitudinal studies 
remain limited, due to methodological challenges in analysing 
longitudinal data of 24h behaviours as a composition, 
which requires accounting for their multilevel structure in addition to the non-Euclidean properties of compositional data.

\textbf{Sleep Architecture.} Sleep architecture comprises total awake time in bed (TWT; sleep onset latency [SOL] 
plus wake after sleep onset [WASO]), light sleep (non-rapid eye movement [NREM] stages 1 and 2), 
slow-wave sleep (SWS; also referred to as NREM stage 3), and Rapid Eye Movement (REM) sleep \parencite{iber2007}.
Sleep architecture data has traditionally been collected over as few as one night spent in a 
sleep laboratory (e.g., clinical assessments, experimental studies).
However, data from multiple nights of sleep offer a more comprehensive profile of an 
individual's sleep, including to characterise both habitual sleep and within-person 
variability of sleep across nights. Assessments of sleep architecture in longitudinal, 
daily studies in naturalistic (at home) settings have become available using ambulatory 
electroencephalographic sleep-monitoring devices. 
Their use in research is also
increasing \parencite{yap2022, spina2023}, resulting in the growth of multilevel sleep architecture data.

Most people have limited sleep opportunity each night. 
Thus, the times spent in different sleep stages are constrained by the total time spent in bed. 
Sleep architecture composition is the distribution of distinct sleep stages within time in bed, 
whereby both the absolute and relative times in different stages are 
informative. When time in bed is fixed, an increase of time in one sleep stage must be proportionally 
countered by a corresponding reduction in time spent in other stages.
Existing evidence supports this notion, showing that combinations of sleep architecture alterations 
can characterise mental disorders better than alterations in one single sleep stage \parencite{baglioni2016}, such as the expression of SWS and consequential overexpression of REM in 
depression \parencite{palagini2013}. However, to our knowledge, research is yet to analyse 
sleep architecture across multiple nights as multilevel compositional data. 
Modelling sleep architecture as a composition could lead to new insights into 
the daily determinants and consequences of sleep.

\subsubsection{Dietary Macronutrients}
Nutrient data are naturally compositional, as they are parts of 
complex food matrices and not consumed in isolation. For example, carbohydrates, fat and 
protein are components of the nutrient composition. Notably, considering nutrients holistically is 
important as an increased intake of one nutrient can influence the absorption or the use of another.
Nutrient are often repeatedly measured over time, thus, are also multilevel.

Nutritional research on diet–disease associations has employed CoDA to model 
the balances of nutritional components \parencite{leite2016, leite2019}, but to our best knowledge, not yet in a multilevel framework.
Conceptualising nutrients as a composition aligns with the notion that 
metabolic dysfunction may not only be due to a deficiency or excess of a
particular nutrient, but may also be due to a loss of balance between
nutrients \parencite{leite2019}. Nutrition is linked with psychological factors.
For example, diet is a modifiable lifestyle factor for the prevention and treatment
of mental disorders \parencite{firth2020}, and nutritional interventions have 
potential to protect or promote psychological well-being \parencite{grajek2022}. 
However, as with sleep and other behaviours, a single day or only an 
average nutrient profile is only an ``snapshot'', whereas assessing and modeling nutrient profiles 
over multiple days using multilevel CoDA has potential to offer new insights 
into short-term, prospective impacts of diet and daily factors that may drive choices around
food intake and nutrient profiles.

\subsubsection{Forced-choice Items}
Ipsative assessments, or forced-choice scales, have been used in questionnaires,
but not commonly due to challenges in analysing the data \parencite{smithson2024}. 
Ipsative data, or other forced-choice data, can be classified as compositional data, 
as the scores in a variable are dependent on other variables which are assessed,
and the sum of the scores obtained over the attributes measured for each respondent
is constant. These data become multilevel when they are repeatedly measured across
respondents or respondents are clustered in different groups
(e.g., children nested within schools, employees nested within companies).

The Occupational Personality Questionnaire is an example of an ipsative inventory.
The questionnaire was originally developed in two versions: a
normative rating scale version and a forced-choice format ipsative scale.
The normative version, while commonly used, is subject to response biases such as
social desirability, halo effects, or impression management \parencite{joubert2013}.
The ipsative version, in contrast, reduces response bias by employing forced-choice items.
Items constructed with an ipsative approach present respondents with options equal in
desirability so they cannot endorse all items, and instead are required to
weigh the relative importance of them \parencite{cunningham1977, bowen2002}.
This ipsative version, was replaced with item response theory to generate normative
scale scores \parencite{joubert2013}, due to challenges in analysing the data.
However, the dependencies due to the compositional nature of ipsative data can be appropriately modelled using CoDA, as explained in the following section.

\subsection{Multilevel Compositional Data on the Simplex}
Detailed structure of single-level compositional data and the relevant data 
transformations have been described previously \parencite{dumuid2018, van2013, smithson2024}. 
We recommend readers who are unfamiliar with CoDA consulting one of those sources first.
Here, we extend the fundamental concepts of compositional data to a multilevel framework.

For $d = 1, \ldots, D$ part composition at  $i = 1, \ldots, I$ time points 
for $j = 1, \ldots, J$ individuals, a multilevel composition is defined as 
a vector of $D$ positive parts that sum to a constant $\kappa$. 
We denote the multilevel composition observed at the $i^{\text{th}}$ time point for the $j^{\text{th}}$ person as

\begin{equation}
\pmb{x}_{ij} = (x_{1ij},x_{2ij},\ldots,x_{Dij}), \hspace{0.1cm}
\text{where} \hspace{0.1cm} \sum_{d=1}^D x_{dij} = \kappa
\label{eq-mcomp}
\end{equation}
Compositions are elements in the $D$-simplex, denoted as $\pazocal{S}^D \subset \mathbb{R}^D$, 
where all $D$-compositional parts are constrained to sum to a constant, $\kappa$.
For example, the time spent in a day dedicated to sleep, physical activity, 
and sedentary behaviour forms a composition represented on the simplex and defined 
by the sum constraint of the 24 hours ($\kappa = 24$). 
Consequently, standard mathematical operations (e.g., addition, multiplication)
are incompatible within the geometry of the simplex because they do not guarantee
that the sum remains $\kappa$ (e.g., $\pazocal{S}^D$ is not closed under addition). We describe some important
properties of the Simplex that are relevant to the analysis of multilevel compositional data in the following.

\textbf{Perturbation.} Perturbation in the simplex 
($\pazocal{S}^D$), or the closure operation applied to the element-wise product,
is the analogous operation to addition in Euclidean space ($\mathbb{R}^{D - 1}$) \parencite{van2013, aitchison1982}. Perturbation of two compositions requires perturbing the relative value of each part
of the composition \parencite{aitchison1982}. This is defined as 

\begin{equation}
\pmb{x}_{ij} \oplus \pmb{x}^{*}_{ij} = \pazocal{C}(
x_{1ij} \pmb{\cdot} x^{*}_{1ij}, 
x_{2ij} \pmb{\cdot} x^{*}_{2ij}, 
\ldots, 
x_{Dij} \pmb{\cdot} x^{*}_{Dij}
)
\end{equation}
where
$$
\pazocal{C}(\pmb{x}_{ij})= \frac{\kappa}{\sum_{d=1}^D x_{dij}}\pmb{x}_{ij}
$$
is the closure operation that normalises the compositional parts of a 
vector $\pmb{x}_{ij}$ sum to the constant $\kappa$ \parencite{aitchison1982}, and
$\pmb{x}_{ij}, \pmb{x^{*}}_{ij} \in \pazocal{S}^D$. Perturbation is an associative and commutative operation; the neutral element is $1_{D} = (1, 1, \ldots, 1)$
and the opposite element is 
$\ominus \hspace{0.1cm} \pmb{x} = \pazocal{C}\left( \frac{1}{x_{1ij}}, \ldots, \frac{1}{x_{Dij}} \right)$.

\textbf{Powering.} The power transformation replaces the product of a vector by a scalar and is defined as the closed powering of the components by a given scalar $\alpha \in \mathbb{R}$
\begin{equation}
\pmb{x}_{ij} \odot \alpha = \pazocal{C}(
x^{\alpha}_{1ij}, x^{\alpha}_{2ij}, \ldots, x^{\alpha}_{Dij})
\end{equation}

\textbf{Inner Product.} The Aitchison inner product of 
$\pmb{x}_{ij}$ and $\pmb{x^{*}}_{ij}$ is defined as
\begin{equation}
{\langle\pmb{x}_{ij}, \pmb{x}^{*}_{ij} \rangle}_a
= 
\sum_{d=1}^{D} \ln\frac{{x}_{dij}}{g(\pmb{x}_{ij})}\ \ln\frac{{x}^{*}_{dij}}{g(\pmb{x}^{*}_{ij})} 
= 
\frac{1}{D} \sum_{d < d'} 
\ln\frac{{x}_{dij}}{{x}_{d'ij}}
\ln\frac{{x}^{*}_{dij}}{{x}^{*}_{d'ij}}
\end{equation}
where $g(\cdot)$ denotes geometric mean of parts.
The subscript $a$ refers to the specific Aitchison geometry operation, in order to distinguish it from the standard inner product used in
$\mathbb{R}$.

\subsection {Log-ratio Approach for Multilevel Compositional Data Analysis}
When modelling data where a subset of the data are compositional, the inclusion of all compositional parts in a 
single analytical model is problematic due to the perfect multi-collinearity between them.
CoDA \parencite{aitchison1982, Pawlowsky2011} 
is a log-ratio analysis paradigm that utilises the relative information contained 
in compositional data. Several transformations exist 
\parencite[for discussions, see][]{dumuid2018, van2013}. A common transformation is  
the isometric log-ratio ($ilr$) \parencite{egozcue2003}. The $ilr$ transformation
preserves the metric properties of the composition and accounts for the dependencies 
between its parts, so that standard statistical methods can be applied to the 
transformed data. The $\text{ilr}()$ function involves transforming the $D-$part composition in the 
simplex ($\pazocal{S}^D$) to a set of $(D-1)-$dimension $ilr$ coordinates in the 
Euclidean space ($\mathbb{R}^{D - 1}$) isometrically (i.e., preserving angles and distances).
Specifically, a $D-$part composition $\pmb{x}_{ij} \in \pazocal{S}^D$ can be re-expressed as
its corresponding set of $D-1$ $ilr$ coordinates using the $\text{ilr}()$ function

\begin{equation}
\text{ilr}(\pmb{x}_{ij}) = \pmb{z}_{ij} = 
(z_{1ij}, z_{2ij},\ldots,z_{(D-1)ij})  \in \mathbb{R}^{D - 1}
\label{eq-milr}
\end{equation}

\subsubsection{Sequential Binary Partition}
As there is not one unique ilr transformation, a valid orthonormal basis needs to be chosen. The isometry from $\pazocal{S}$ to $\mathbb{R}$ 
is commonly constructed using a sequential binary partition (SBP),
a $D \times (D-1)$ matrix that maps the $D$ compositional parts and 
their membership in the $(D-1)$ $ilr$ coordinates \parencite{egozcue2005}.
A SBP is obtained by first partitioning the compositional parts into two non-empty sets, 
where one set corresponds to the first $ilr$ coordinate's numerator (coded as + 1) and 
the other set corresponds to the first $ilr$ coordinate's denominator (coded as -1), and 
where applicable, compositional part(s) uninvolved in the $ilr$ are coded as 0.
Using this principle, each of the previously constructed sets are recursively partitioned 
into two non-empty sets until no further partitions of the subcompositional 
parts are possible (after $D - 1$ steps).
The $ilr$ coordinates can be interpreted as the log-ratio of the subcomposition 
in the numerator in relation to the subcomposition in the denominator.
Table \ref{tab-sbp} gives an example of a complete SBP for a five-part composition $\pmb{x}_{ij} = (x_{1ij}, x_{2ij}, x_{3ij}, x_{4ij}, x_{5ij})$.
Here, the first binary partition separates two groups of parts
$[x_{1ij}, x_{2ij}]$ coded as + 1 and $[x_{3ij}, x_{4ij}, x_{5ij}]$ coded as - 1. The second partition is made of two groups of parts
$[x_{1ij}]$ coded as + 1, $[x_{2ij}]$ coded as - 1, with 
$[x_{3ij}, x_{4ij}, x_{5ij}]$ coded as 0.
Note the partitions can only be made on parts that have not been separated by grouping in the previous partitions.
The SBP ends at step (D-1), that is 4 in this example. 
Although the order of parts in composition might be mathematically arbitrary,
the order of SBP can be constructed to be interpretable.
For example, we can order the parts to ensure that the 
SBP forms conceptually meaningful contrasts
(e.g., time spent in sleeping behaviours all relative to waking behaviours).
Even when the $ilr$ coordinates resulting from any given SBP may be difficult
to interpret, it is possible to rely on post-hoc substitution analysis for 
interpretation, which is introduced later. Using substitution analysis, the choice of a SBP (or other valid ilr bases) used becomes irrelevant as the 
substitution analysis can evaluate all possible pairwise reallocations across compositional parts.

\begin{table*}[!htbp]
  \caption{Example Sequential Binary Partition of A Five-Part Composition.}
  \centering
    {\renewcommand{\arraystretch}{1.25}% for the vertical padding
    \setlength\tabcolsep{0pt}
    \begin{tabular*}{\linewidth}{@{\extracolsep{\fill}} lrrrrr}
\toprule
Partition order & $x_{1ij}$ & $x_{2ij}$ & $x_{3ij}$ & $x_{4ij}$ & $x_{5ij}$ \\
\midrule
1 & +1 & +1 & -1 & -1 & -1 \\
2 & +1 & -1 &  0 &  0 &  0 \\
3 & 0  & 0  & +1 & -1 & -1 \\
4 & 0  & 0  &  0 & +1 & -1 \\
\bottomrule
  \end{tabular*}}
\label{tab-sbp}
\end{table*}

\subsubsection{Orthonormal Basis of a Partition}
The SBP matrix provides an orthonormal basis of $\pazocal{S}^{D}$ 
and allows the constructions of coordinates that are the balances between the groups of parts separated in each step of a binary partition.
For example, 
in the $k$-order binary partition, 
we may separate $r$ parts 
$x_{(d + 1)ij}, \ldots, x_{(d + r)ij}$ from $s$ parts
$x_{(d + r + 1)ij}, \ldots, x_{(d + r + s)ij}$.
We denote the remaining parts in the composition 
that are not involved in the partition as 
$x_{1ij}, \ldots, x_{dij}$ ($d$ parts) 
and 
$x_{(d + r + s + 1)ij}, \ldots, x_{Dij}$ ($d'$ parts).
Without loss of generality, $D = d + r + s + d'$ and $k \leq D - r - s + 1$, and $d$ and $d'$ can be zero.
The \textit{balancing element} associated with the $k$-order binary partition
$\pmb{e}_{k}$ is defined in \textcite{egozcue2005} as

\begin{equation}
\pmb{e}_{k} = \pazocal{C} 
\begin{bmatrix}
\exp \left( 
\underbrace{0, \ldots, 0}_{d  \text{ elements}},
\underbrace{a, \ldots, a}_{r  \text{ elements}},
\underbrace{b, \ldots, b}_{s  \text{ elements}},
\underbrace{0, \ldots, 0}_{d' \text{ elements}}
\right)
\end{bmatrix}
\label{e_k}
\end{equation}
where
$$
a = \sqrt{\frac{{s_k}}{r_k(r_k + s_k)}}
\hspace{0.1cm} \text{and} \hspace{0.1cm} 
b = -\sqrt{\frac{r_k}{s_k(r_k + s_k)}}
$$
For each SBP, the $D-1$ balancing elements 
uniquely define an associated orthonormal basis. 
For example, the complete basis elements associated with the SBP of Table \ref{tab-sbp} are
\begin{equation}
\begin{aligned}
e_1 &=
\pazocal{C} 
\begin{bmatrix}
\exp \left( 
\sqrt{\frac{3}{2 \cdot 5}}, 
\sqrt{\frac{3}{2 \cdot 5}}, 
-\sqrt{\frac{2}{3 \cdot 5}}, 
-\sqrt{\frac{2}{3 \cdot 5}}, 
-\sqrt{\frac{2}{3 \cdot 5}}
\right)
\end{bmatrix} \\
e_2 &= 
\pazocal{C} 
\begin{bmatrix}
\exp \left(
\sqrt{\frac{1}{1 \cdot 2}}, 
-\sqrt{\frac{1}{1 \cdot 2}}, 
0, 0, 0
\right)
\end{bmatrix} \\
e_3 &=
\pazocal{C} 
\begin{bmatrix}
\exp \left(
0, 0, 
\sqrt{\frac{2}{1 \cdot 3}}, 
-\sqrt{\frac{1}{2 \cdot 3}}, 
-\sqrt{\frac{1}{2 \cdot 3}}
\right)
\end{bmatrix} \\
e_4 &=
\pazocal{C} 
\begin{bmatrix}
\exp \left(
0, 0, 0, 
\sqrt{\frac{1}{1 \cdot 2}}, 
-\sqrt{\frac{1}{1 \cdot 2}}
\right)
\end{bmatrix}
\end{aligned}
\label{basis}
\end{equation}

\subsubsection{The Isometric Log-ratio Coordinates}
We denote $z_{kij}$ as
the $k^{^{\text{th}}}$ ($k = 1,2,\ldots, D-1$) $ilr$ coordinate
observed at time point $i$ for individual $j$ and can be shown to be the coordinate of $\pmb{x}_{ij}$ 
with respect to the balancing elements 
$\pmb{e}_{k}$ \parencite{egozcue2003},
\begin{equation}
\begin{aligned}
z_{kij} 
&=
{\langle \pmb{x}_{ij}, \pmb{e}_{k} \rangle}_a \\[6pt]
&=
\sqrt{\frac{r_k s_k}{r_k + s_k}}
\ln
\begin{bmatrix}
\dfrac{g\left(x_{(d + 1)ij}, \ldots, x_{(d + r)ij}\right)}{g\left(x_{(d + r + 1)ij}, \ldots, x_{(d + r + s)ij}\right)}
\end{bmatrix} \\[8pt]
&=
\ln
\begin{bmatrix}
\dfrac{
{\left(x_{(d + 1)ij}     \ldots x_{(d + r)ij}\right)}^{\sqrt{ \sfrac{s_k}{r_k(r_k + s_k)}}}}{
{\left(x_{(d + r + 1)ij} \ldots x_{(d + r + s)ij}\right)}^{\sqrt{\sfrac{r_k}{s_k(r_k + s_k)}}}}
\end{bmatrix} 
\end{aligned}
\end{equation}
where $g(\cdot)$ refers the geometric mean of the arguments. 
The r parts $(x_{(d + 1)ij}, \ldots, x_{(d + r)ij})$ in the first group are coded as +1 and placed in the numerator, and the s parts $(x_{(d + r + 1)ij}, \ldots, x_{(d + r + s)ij})$ in the second group are coded as -1 and placed in the denominator. The coordinates corresponding to the basis \ref{basis} are
\begin{equation}
\begin{aligned}
z_{1ij} &= 
\ln
\begin{bmatrix}
\dfrac{
{(x_{1ij} x_{2ij})}^{\sqrt{\sfrac{3}{10}}}}{
{(x_{3ij} x_{4ij} x_{5ij})}^{\sqrt{\sfrac{2}{15}}}}
\end{bmatrix} \\[4pt]
z_{2ij} &= 
\ln
\begin{bmatrix}
\dfrac{
{(x_{1ij})}^{\sqrt{\sfrac{1}{2}}}}{
{(x_{2ij})}^{\sqrt{\sfrac{1}{2}}}}
\end{bmatrix} \\[4pt]
z_{3ij} &= 
\ln
\begin{bmatrix}
\dfrac{
{(x_{3ij})}^{\sqrt{\sfrac{2}{3}}}}{
{(x_{4ij} x_{5ij})}^{\sqrt{\sfrac{1}{6}}}}
\end{bmatrix} \\[4pt]
z_{4ij} &= 
\ln
\begin{bmatrix}
\dfrac{
{(x_{4ij})}^{\sqrt{\sfrac{1}{2}}}}{
{(x_{5ij})}^{\sqrt{\sfrac{1}{2}}}}
\end{bmatrix}
\end{aligned}
\label{coords}
\end{equation}

The main property of the representation of compositions by their coordinates with respect to an orthonormal basis is that the Aitchison geometry of compositions in the simplex $\pazocal{S}^D$ is reduced to the ordinary Euclidean geometry in $\mathbb{R}^{D - 1}$ for their coordinates.
For example,
\begin{equation}
\text{ilr}(\pmb{x}_{ij} \oplus \pmb{x}^{*}_{ij}) = \pmb{z}_{ij} + \pmb{z}^{*}_{ij}, \quad 
\text{ilr}(\alpha \odot \pmb{x}_{ij}) 
= \alpha \pmb{\cdot} \pmb{z}_{ij}
\end{equation}
Importantly, the $ilr$ coordinates are linearly independent
multivariate real values  \parencite{mateu2011}.
Therefore, once the multilevel composition has been re-expressed as a set of 
corresponding $ilr$ coordinates, they can be entered into standard statistical models,
such as multilevel models. The ilr transformation function is injective and invertible.
That is, the $ilr$ coordinates can be back-transformed via their $1 - 1$ relationship 
to the original composition \parencite{egozcue2003} using

\begin{equation}
\pmb{x}_{ij} = \text{ilr}^{-1}(\pmb{z}_{ij}) = \bigoplus_{k=1}^{D-1} (z_{kij} \odot \pmb{e}_{k})
\end{equation}
where $\bigoplus$ stands for repeated perturbation.
The inverse ilr transformation is convenient, as even the best efforts to construct $ilr$ coordinates based on a SBP are typically less intuitive and interpretable than the estimates of 
the original composition (e.g., minutes spent in sleep, physical activity, and sedentary).
We later explain how Bayesian statistics provide a convenient framework for this inverse transformation in the interpretation of results from multilevel compositional data. We also discuss the interpretation of $ilr$ coordinates in specific real data application. 

\subsection{Disaggregating Levels of Effects}
In this section, we discuss the properties of multilevel compositional data in the context of 
longitudinal studies, specifically in a two-level data hierarchy
(e.g., daily observations nested within people). 
Although we focus on longitudinal data here, the same principles can be applied to distinguish 
effects at different levels of analysis for multilevel data with an arbitrary number of levels. 

When compositional data (e.g., behaviours, diet) are repeated measures on multiple people,
these data contain two sources of variability:
between-person (i.e., differences between individuals) and within-person 
(i.e., changes within individuals).
Recommendations for multilevel models in these cases are to use
person-mean centering to explicitly separate associations that exist 
between people versus those that exist within people \parencite{wang2015}.
Studying these two unique processes open up an avenue to investigate
not only how people with different compositions may vary, 
but also how fluctuations around an individual's own typical 
composition may be associated with outcomes.
Next, we show how the concepts of person-mean centering 
can be applied to multilevel compositional data.

For a $D$-part multilevel composition in $\pazocal{S}^{D}$, the $d^{\text{th}}$ ($d = 1, 2,\ldots, D$) part is the product of its between and within levels, denoted as
\begin{equation}
\pmb{x}_{dij} = x{^{(b)}_{d \pmb{\cdot} j}} \pmb{\cdot} x{^{(w)}_{dij}}
, 
\hspace{0.2cm} k = 1, 2, \ldots, D - 1
\end{equation}
where
\begin{itemize}
    \item $x{^{(b)}_{d \pmb{\cdot} j}}$ is the person-specific mean of the $d^{\text{th}}$ compositional part over time,
    which contains only between-person variance and no within-person variance.
    The subscript $\pmb{\cdot}j$ denotes the average across $i$ observations for the individual $j$ and superscript $^{(b)}$ denotes the $between$ level of the compositional parts.
    \item $x{^{(w)}_{dij}}$ is the time-specific deviation (at time $i^{\text{th}}$) of the $d^{\text{th}}$ compositional part from the person $j$ specific mean (i.e., compositional mean-centered deviate), which has within-person variance and no between-person variance. The superscript $^{(w)}$ denotes the $within$ level of the composition parts.
\end{itemize}
The complete multilevel composition (Equation. \ref{eq-mcomp}) can be re-expressed as its between- and within-person parts as

\begin{equation}
\begin{aligned}
\pmb{x}_{ij} 
&= 
\pazocal{C} (x_{1ij},x_{2ij},\ldots,x_{Dij}) \\
&=
\pazocal{C} 
\begin{pmatrix}
x{^{(b)}_{1 \pmb{\cdot} j}} \pmb{\cdot} x{^{(w)}_{1ij}}, 
x{^{(b)}_{2 \pmb{\cdot} j}} \pmb{\cdot} x{^{(w)}_{2ij}}, 
\ldots, 
x{^{(b)}_{D \pmb{\cdot} j}} \pmb{\cdot} x{^{(w)}_{Dij}}  
\end{pmatrix}
\\
&=
\pmb{x}{^{(b)}_{\pmb{\cdot} j}} \oplus \pmb{x}{^{(w)}_{ij}} \\
\end{aligned}
\end{equation}
with $\oplus$ being the perturbation operation on the simplex, and
$\pazocal{C}$ being the closure operation.
The between- and within-person subcompositions are themselves compositions
\begin{equation}
\begin{aligned}
\pmb{x}{^{(b)}_{ \pmb{\cdot} j}}    
&= 
\pazocal{C} \left(
x{^{(b)}_{1 \pmb{\cdot} j}},
x{^{(b)}_{2 \pmb{\cdot} j}},
\ldots,
x{^{(b)}_{D \pmb{\cdot} j}} \right) \text{ and } \\
\pmb{x}{^{(w)}_{ij}}          
&= 
\pazocal{C} \left(x{^{(w)}_{1ij}}, x{^{(w)}_{2ij}}, \ldots, x{^{(w)}_{Dij}}\right)
\end{aligned}
\end{equation}

As the $ilr$ coordinates exist in the Euclidean space $\mathbb{R}^{D - 1}$,
the decomposition of the $(D - 1)-$dimension $ilr$ coordinates $\pmb{z}_{ij}$ (Equation. \ref{eq-milr})
can be achieved using the usual addition operation, that is
\begin{equation}
\begin{aligned}
\pmb{z}_{ij}
&=
(z_{1ij}, z_{2ij},\ldots,z_{(D-1)ij}) \\
&=
\left( z{^{(b)}_{1    \pmb{\cdot} j}} + z{^{(w)}_{1ij}}, 
z{^{(b)}_{2     \pmb{\cdot} j}} + z{^{(w)}_{2ij}}, 
\ldots, 
z{^{(b)}_{(D-1) \pmb{\cdot} j}} + z{^{(w)}_{(D-1)ij}} \right)
\\
&=
\pmb{z}{^{(b)}_{\pmb{\cdot} j}} + \pmb{z}{^{(w)}_{ij}}
\end{aligned}
\end{equation}
in which superscript $^{(b)}$ and $^{(w)}$ also denote the between and within levels of 
the $ilr$ coordinates.

Here we have focused on longitudinal data common in psychology, that is repeated measures are nested within people as has been done in previous papers \parencite{wang2015}.
However, the same principles can be applied to distinguish different levels of effects in hierarchical data, that is when observations are clustered (e.g., individuals nested within groups). Classic examples of hierarchical data include children within schools and patients within hospitals. In these cases, the same steps can be applied but the interpretation of ``within person'' $ilr$ coordinates would instead be the $ilr$ coordinates at the lowest level (e.g., children) and the ``between person'' $ilr$ coordinates reflect the higher level (e.g., schools).
The equations outlined here to separate the effects of compositional variables at different levels of analysis only work well with two-level data structure, wherein between-cluster level is cluster-mean at level 2, and within-cluster level is the mean-centered deviate at level 1. 
Separating effects across more-than-two-level data hierarchy is outside the scope of this current work. Likewise, disaggregating results for cross-classified data structures (e.g., children nested within schools and neighbourhoods, where there are two clusters that are not themselves nested)
remains to be developed. Recent research has made recommendations for disaggregating cross-classified multilevel models for non-compositional data \parencite{guo2024}; the same strategy could in principle be translated to compositional data in future work.
Presently, for multilevel compositional data with more than two levels or a cross-classified structure, our recommendation is to keep the data at the aggregate level (i.e., not separated by between and within-cluster effects), and exercise care in the interpretation
\parencite[see][for a discussion on between-cluster and within-cluster inferences]{Curran2011}.

\section{Multilevel Modelling using Bayesian Inference}
\subsection{Bayesian Approach to Multilevel Modelling}
Our exposition of Bayesian inference will be kept to a minimum, 
given the rich and growing literature that 
offers methodological guidance on Bayesian analyses, including comprehensive coverage from beginning through advanced topics \parencite{kruschke2014, mcelreath2018, gelman2013bayesian}.
Here we discuss our Bayesian perspectives on the proposed method, 
focusing on its computational flexibility when estimating complex models, including multilevel models, and performing post-hoc analyses. 
We also briefly explain the prior specification required for this approach.

\subsubsection{Computational Flexibility}
The Bayesian approach offers computational flexibility for multilevel CoDA. 
Bayesian statistics considers each parameter of a model a random variable (as opposed to a 
frequentist framework where parameter values are unknown constants), 
which requires the explicit use of probability to model the uncertainty in prediction.
Consequently, all Bayesian models by default come with the probability distribution of parameters, 
allowing for the point summary (e.g., a posterior mean, median, or mode) and 
uncertainty (e.g., standard errors, credibility intervals) to be directly and intuitively calculated. 
This is particularly relevant for multilevel CoDA, as it involves log-ratio transformations, which 
benefits from post-hoc analyses to aid interpretation of results. For example, the estimation procedure 
of models with compositional outcomes may include transforming compositions into $ilr$ coordinates, 
estimating the multilevel models, and back-transforming the $ilr$ coordinates to the original compositions 
to obtain straightforward results. For example, the number of minutes or hours spent in each behaviour on 
weekdays versus weekend, or the difference in minutes between weekdays and weekends for each behaviour, 
instead of estimated $ilr$ coordinate differences. Similarly, when estimating models with compositional 
predictors, we are often interested in the expected difference in the outcome when a fixed amount of the 
composition is reallocated from one compositional part to another (e.g., estimated differences in 
depressive symptoms when reallocating 30 minutes to physical activity at the expense of sedentary 
behaviour). These estimates and their inferences can be calculated using a series of post-hoc predictions 
referred to as substitution analysis in the CoDA literature. 
A challenge with substitution analysis in a frequentist framework is that transforming predictions to 
the original scale, often more interpretable, involves non-linear transformations and then calculating differences. Appropriately calculating uncertainty (e.g., confidence intervals) in a 
frequentist framework typically involves bootstrapping. Under the Bayesian paradigm, we can use 
the posterior distributions, which intuitively without adding coding enables 
accurate estimates and credible intervals to be calculated.
We will later discuss this analysis in more details and explain how it can 
substantially enhance the interpretation and communication of results.

It is important to acknowledge that Bayesian sampling algorithms, such as 
Markov chain Monte Carlo (MCMC), often require longer run time than frequentist estimation 
methods, such as maximum likelihood (ML) or restricted ML. However, we believe that in most cases,
post-hoc substitution analyses are desirable and that quantifying uncertainty in these substitutions
is an important part of multilevel CoDA. Bootstrapping a frequentist model would also increase 
the total run time and requires additional code implementation.
Thus, we find the straightforward Bayesian model setup and estimation 
outweigh the trade-off in computational resources. The rapid increase in computational resources 
and user-friendly software have also facilitated accessibility to Bayesian analysis, including \textbf{Stan} \parencite{carpenter2017, stan2023}, the \textbf{R} package \textit{brms} \parencite{brms, burkner2017} for Bayesian modelling generally, and the \textbf{R} package \textit{multilevelcoda} for multilevel CoDA particularly \parencite{multilevelcoda}.

\subsubsection{Exchangeability and Multilevel Modelling}
Models emerge in a Bayesian context under the principle of exchangeability.
Exchangeability refers to the invariance of the Bayesian model to any permutation of the parameters, 
which corresponds to the belief that the order of the observations is irrelevant.
A simple example is tossing a coin twice, where we assume the probability of getting 
one heads is unaffected by whether it appears in the first or the second toss.
The de Finetti's Representation Theorem, a probability theory underpinning Bayesian statistics, 
proves that data that are exchangeable come from the same unknown population. 
That means exchangeability of the 
individual (lower-level) units is achieved by conditioning (i.e., partially pooled)
on the population (higher-level) units \parencite{gelman2013bayesian}.
Thus, Bayesian statistics treat data as conditionally-independent and model the dependency between levels of units. 
This is mathematically equivalent to assuming a hierarchical structure of data \parencite{bernardo2009}, making Bayesian statistics a useful and convenient framework for multilevel models.

\subsubsection{Prior Specifications}
Bayesian inference requires the specifications of prior distributions,
which reflect knowledge about the relative plausibility of
parameter values before data collection.  
Once the data are observed, 
the knowledge in the prior distribution is incorporated to compute the posterior distribution.
That is, the posterior distribution represents the updated 
knowledge from the prior distribution about a parameter in a model given the data, which is mathematically described by Bayes' theorem as

\begin{equation}
\underbrace{p(\theta | y)}_{\text{Posterior}} = \overbrace{p(\theta)}^{\text{Prior}} 
\hspace{0.1cm} \times \hspace{0.1cm}
\frac{\overbrace{p(y | \theta)}^{\text{Likelihood}}}{\underbrace{p(y)}_{\text{Marginal Likelihood}}}
\end{equation}
where $p(\theta | y)$ is the posterior distribution of the parameter given the data $y$, 
$p(y | \theta)$ is the likelihood of the data given the parameter values $\theta$, 
$p(\theta)$ is the prior distribution of the parameter, and 
$p(y)$ is the marginal likelihood of the data. 

Although the shape of the prior distribution influences the shape of the posterior distribution,
researchers can either assume a subjective or a default ``objective'' prior distribution, 
depending on the Bayesian perspectives they adopt \parencite{levy2023}.
A subjective prior distribution reflects the expectations of a researcher on the model parameters.
Information to be incorporated in the prior distribution can be obtained from 
practical or theoretical considerations, or derived from findings of previous studies, 
or elicited from expert knowledge \parencite{stefan2022, van2014}.
Subjective, informative prior distributions have the additional advantage 
that evidence for or against a model can typically accrue faster than when 
default priors are used \parencite{stefan2019, stefan2022}.

Conversely, a default prior distribution is not tailored to reflect a specific prior belief.
Instead, default priors are typically designed with the goal that, when the data is highly informative, the prior is sufficiently diffuse and the likelihood dominates the posterior
\begin{equation}
\underbrace{p(\theta | y)}_{\text{Posterior}} \approx
\frac{\overbrace{p(y | \theta)}^{\text{Likelihood}}}{\underbrace{p(y)}_{\text{Marginal Likelihood}}}
\end{equation}

\begin{figure*}[!htbp]
  \centering
  \caption{Examples of the prior and likelihood combining to influence the posterior \parencite{schad2021}. 
  When data constrain the parameters through the likelihood, then a default, flat prior is sufficient to obtain a concentrated posterior (A). When the data does not sufficiently constrain the parameters through the likelihood, then using a flat prior leaves the posterior diffuse (B), whereas using a (weakly) informative prior helps constrain the posterior to reasonable values (C).}
  \includegraphics[width=\columnwidth]{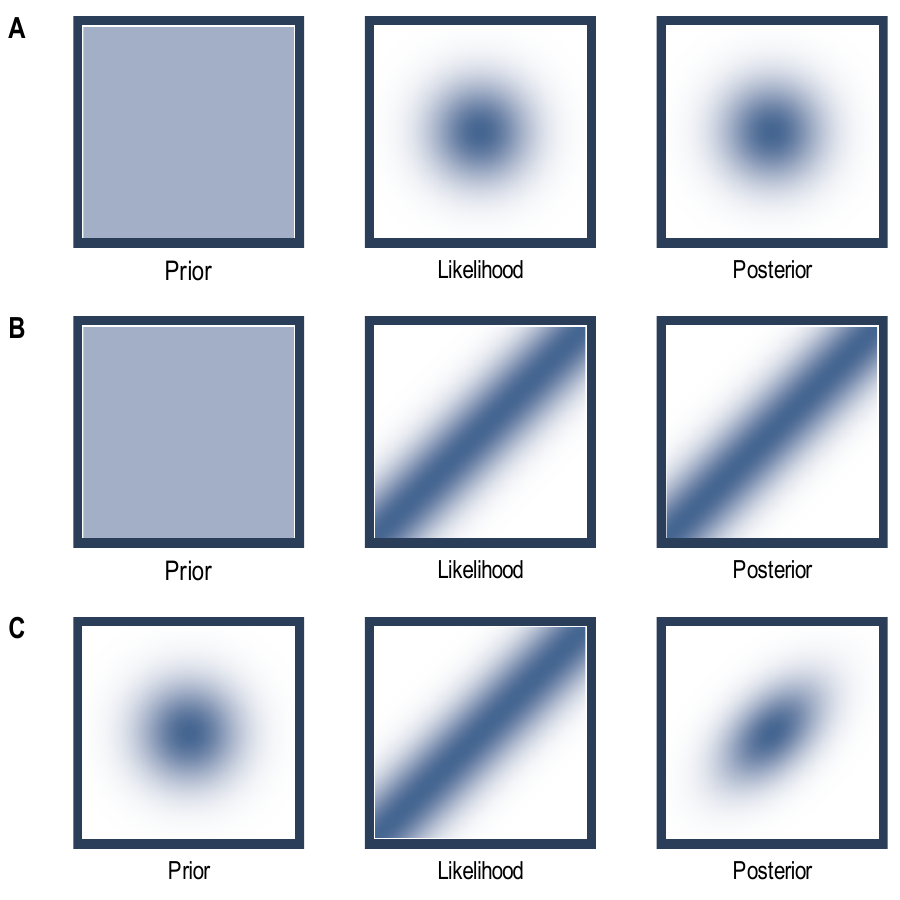} \\
\label{fig-prior}
\end{figure*}

Figure \ref{fig-prior} provides examples of the influences of prior and likelihood on the posterior \parencite{schad2021}. Here, a default ``flat'' prior leaves the posteriors looking like the likelihood, regardless of whether the data constrain the parameters through the likelihood (Figure \ref{fig-prior}A and \ref{fig-prior}B). The minimal influence of the prior on the posterior is the outcome that many default priors aim for. Choosing good priors is particularly relevant in situations where the likelihood is weakly informed (Figure \ref{fig-prior}C). This often occurs when a maximal model is fitted to a small dataset that does not constrain estimation of all the variance and covariance parameters of the random effects, resulting in convergence problems in frequentist methods. In such situation, using a more subjective, informative prior, rather than a flat prior, to suppress extreme but not impossible parameter values, may allow fitting and interpreting models that cannot be validly estimated using frequentist tools.
Nevertheless, as no prior can be universally applicable, 
prior sensitivity should ideally be checked. 

In the specific case of multilevel CoDA, 
prior specification is challenging because it is an emerging method.
There is a lack of data analysed using multilevel CoDA that could be used to
inform priors. For the same reason, there is limited methodological and field experts that could provide comprehensive prior knowledge.
Given the complex distribution of compositional data and their corresponding $ilr$ 
coordinates, efforts to specify priors are particularly open to concerns of subjectivity.
Therefore, researchers using this method may wish to employ default priors and give primacy to the data.
Alternatively, they may adopt priors as a secondary supporting role 
while conducting prior and likelihood sensitivity analyses to determine how influential prior choices are on the parameter estimates. 
Finally, we note that our discussion does not preclude 
other Bayesian perspectives \parencite[for a comprehensive review of perspectives on 
Bayesian inference, see][]{levy2023} or prior specifications, but rather serves as 
a starting point for the emerging method of multilevel CoDA.
As multilevel CoDA is more widely adopted and empirical evidence is accumulated,
further guidance on prior specification for multilevel CoDA  
will be a valuable contribution to the community.

\subsection{Bayesian Multilevel Model Description}
\subsubsection{A General Description}
The core of every multilevel model is the prediction of the response $\pmb{y}$ through the linear combination
$\eta$ of predictors transformed by the inverse link function $g^{-1}$ 
assuming a certain distribution $D$ for $\pmb{y}$
\begin{equation}
    \pmb{y} \sim D(g^{-1}(\eta), \theta)
\end{equation}
where the parameter $\theta$ describes additional family specific parameters
that typically do not vary across data points, such as the standard deviation
$\sigma$ in normal models or the shape $\gamma$ in Gamma or negative binomial models.
The linear predictor can generally be written as
\begin{equation}
    \eta = \pmb{X}\beta + \pmb{Z}u
\end{equation}
where $\beta$ and $u$ are
population-level (fixed) and group-level (random) effects, respectively, and
$\pmb{X}$ and $\pmb{Z}$ are the corresponding design matrices.
The outcome $\pmb{y}$, $\pmb{X}$ and $\pmb{Z}$ make up the data, whereas
$\beta$ and $u$ are the model parameters being estimated.

As a starting point, we denote an simple multilevel model predicting an outcome $y_{ij}$
observed at time point $i$ for individual $j$ by a linear combination of an
intercept $\beta_{0j}$ that is allowed to vary according to the individual $j$
and a slope $\beta_1$ that quantifies the influence of a predictor $x_{ij}$ as
\begin{equation}
\begin{aligned}
    y_{ij}           &=    \mu_{ij} + \varepsilon_{ij} \\
    \eta_{ij}        &=  \beta_{0j} + \beta_1 x_{ij}, 
    \hspace{0.2cm} \mu_{ij} = g^{-1}(\eta_{ij}) \\
   \beta_{0j}        &=    \gamma_{0} + u_{0j}                         \\
   u_{0j}            &\sim \text{Normal}(0, \sigma^{2}_{u})            \\
   \varepsilon_{ij}  &\sim \text{Normal}(0, \sigma^{2}_{\varepsilon}) \\
\end{aligned}
\end{equation}
In the case of a linear, normally distributed outcome,
the link function $g$ is the identity function, thus, the inverse link 
$g^{-1}$ is also the identity function, giving $\mu_{ij} = \eta_{ij}$.
The above multilevel model is thus strictly equivalent to the following Bayesian multilevel model
\begin{equation}
\begin{aligned}
y_{ij}     &\sim \text{Normal}(\mu_{ij}, \sigma^{2}_{\varepsilon}) \\
\mu_{ij}   &=    \beta_{0j} + \beta_1 x_{ij}                       \\
\beta_{0j} &\sim \text{Normal}(\gamma_0, \sigma^{2}_{u})          \\
\end{aligned}
\label{eq-b0j}
\end{equation}
We can add a group-level slope according to $j$ as follows
\begin{equation}
\begin{aligned}
y_{ij}     &\sim \text{Normal}(\mu_{ij}, \sigma^{2}_{\varepsilon}) \\
\mu_{ij}   &= \beta_{0j} + \beta_{1j} x_{ij}                       \\
  \begin{bmatrix} 
    \beta_{0j} \\
    \beta_{1j}  
  \end{bmatrix} 
  & \sim \text{MVNormal}(\pmb{\gamma}, \pmb{\Sigma_u}) \\
\end{aligned}
\label{eq-b0j-b1j}
\end{equation}

\subsubsection{Multilevel Model with Compositional Variables}
The same multilevel modelling principles can be applied to build multilevel models 
with compositional variables. Once the multilevel composition is 
re-expressed as a set of $ilr$ coordinates, they are essentially 
multivariate variables that can be entered into any standard multilevel models, 
as outcomes, predictors, or both. 
For simplicity, we describe a simple multilevel model with compositional predictor
with a group-level intercept. This model will be used in both the real 
data study and simulation study.

Consider a continuous, normally distributed outcome variable 
observed at time point $i$ for individual $j$ as $y_{ij}$ predicted by a compositional predictor
$\pmb{x}_{ij}$ expressed as a set of $ilr$ coordinates $\pmb{z}_{ij}$,
the linear multilevel model of $y_{ij}$ from Equation \ref{eq-b0j} can be extended to
\begin{equation}
\begin{aligned}
    y_{ij}     &\sim \text{Normal}(\mu_{ij}, \sigma^{2}_{\varepsilon})      \\
    \mu_{ij}   &= 
    \beta_{0j} +
    \sum_{k = 1}^{D-1}\beta_k            z_{kij} \\
    \beta_{0j} &\sim \text{Normal}(\gamma_0, \sigma^{2}_{u})
\end{aligned}
\end{equation}
with $z_{kij}$ being the individual $k^{\text{th}}$ ($k = 1,2,\ldots, D-1$) $ilr$ coordinate
observed at time point $i$ for individual $j$.
This model can be expanded to include both the between- and within-person subcompositions 
($\pmb{x}{^{(b)}_{\pmb{\cdot} j}}$ and $\pmb{x}{^{(w)}_{ij}}$), 
expressed as two sets of $ilr$ coordinates ($\pmb{z}{^{(b)}_{\pmb{\cdot} j}}$ and $ \pmb{z}{^{(w)}_{ij}}$), that is
\begin{equation}
\begin{aligned}
    y_{ij}     &\sim \text{Normal}(\mu_{ij}, \sigma^{2}_{\varepsilon})      \\[3pt]
    \mu_{ij}   &= 
    \beta_{0j} +
    \overbrace{\sum_{k = 1}^{D-1}\beta_k            z^{(b)}_{k \pmb{\cdot} j}}^{\text{between}} +
    \underbrace{\sum_{k = 1}^{D-1}\beta_{(k + D - 1)}z^{(w)}_{kij}}_{\text{within}}            \\
    \beta_{0j} &\sim \text{Normal}(\gamma_0, \sigma^{2}_{u}) \\
\end{aligned}
\label{eq-mlm}
\end{equation}
The individual $k^{\text{th}}$ between- and within-person $ilr$ coordinates
are $z^{(b)}_{k \pmb{\cdot} j}$ and $z^{(w)}_{kij}$, with 
the subscripts denoting that the between subcomposition is unique to individual $j$ 
and the within subcomposition is unique to time $i$ for individual $j$.
All $\pmb{z}{^{(b)}_{\pmb{\cdot} j}}$ and $\pmb{z}^{(w)}_{kij}$ are included here as population-level effects, 
however, the $\pmb{z}^{(w)}_{kij}$ can be allowed to vary (added as group-level slopes) to model their 
group-level effects if desired.
The between- and within-person effects of the $ilr$ coordinates are $\beta_k$ and $\beta_{k + D - 1}$.
Because each $ilr$ coordinate is decomposed into its between- and within-person subcompositions,
for $D-1$ $ilr$ coordinates, the number of corresponding $\beta$s for them in the model is $2(D-1)$. Further population- and/or group-level covariates are not included here but can easily be incorporated as required using principles described in Equation \ref{eq-b0j-b1j}.

\section{Bayesian Multilevel Compositional Substitution Analysis}
\subsection{Overview of Compositional Substitution Analysis}
When using multilevel models (Equation \ref{eq-mlm}) to examine 
the association between a compositional predictor and an outcome variable, 
researchers are often interested in which parts of the original composition 
are important to the outcome, by quantifying the change in outcome associated 
with a meaningful change in compositional parts.
This information can not be directly obtained from the estimates of the individual $ilr$ coordinates. 
The coefficients for the $D-1$ $ilr$ coordinates can be back-transformed to the 
composition in the simplex to produce $D$ compositional coefficients.
Compositional coefficients are centred around the compositional zero (a vector of $1/D$s);
a coefficient greater than 1/D suggests a positive change in outcome when the corresponding 
compositional part is relatively increased. As the net change in the outcome, however, 
depends on which other compositional parts are decreased to compensate, 
the compositional coefficients cannot be interpreted in isolation.
To facilitate the interpretation of these models, compositional substitution analysis 
is a post-hoc analysis that examines the expected difference in an outcome when a
fixed unit $t$ of the composition is reallocated from one compositional part to another,
while the other parts remain fixed \parencite{dumuid2019}. Because compositions can be closed
(i.e., collectively sum) to a meaningful amount (e.g., daily behaviours summing to 24 hours or 1440 minutes),
the value for each compositional part corresponds to an absolute amount (e.g., minutes/day spent in that part).
This compositional substitution analysis allows us to investigate how an outcome is associated with the reallocation of a
raw unit from one part of the composition to another.

In behavioural epidemiology, the compositional substitution analysis \parencite{dumuid2018, dumuid2019}
has enabled the investigation of how reallocations from one behaviour
(e.g., minutes/day in sleep, physical activity, and sedentary) to another, while keeping the total
time (e.g., 24 hours) fixed, are associated with physical, mental, and cognitive health outcomes
\parencite{janssen2020, grgic2018, miatke2023}.
In psychological research, there is relatively less uptake of CoDA and compositional
substitution analysis. There remains limited knowledge surrounding how specific reallocations 
of time use (e.g., daily behaviours) or in personality from ipsative tests are associated with 
psychological outcomes. For example, despite the growing evidence from EMA studies 
supporting the independent associations 
between daily behaviours (sleep, physical activity, and sedentary behaviour) and 
emotional experiences and cognitive processes \parencite{hartson2023, shen2022},
there is uncertainty about how daily reallocation of time across behaviours are associated with these phenomena.
The development of a theoretical framework and statistical software that
enables compositional substitution analysis in a multilevel framework
could facilitate more conceptually and analytically meaningful analyses 
using the increasingly available multilevel compositional data.

\subsection{Multilevel Compositional Substitution Analysis}
\subsubsection{Prediction of A Composition}
We extend the compositional substitution analysis \parencite{dumuid2019} to the multilevel framework.
The reallocation of compositional parts, that is, when a fixed unit $t$ of the composition is reallocated from one compositional part to another, while keeping the other parts constant, can be calculated relative to a a starting composition, which we refer to as the reference composition.
Table \ref{tab-submodel} summarises the steps to conduct this analysis using any reference composition 
(e.g., empirical composition based on the sample's mean, theoretical composition based on research question).
A common reference composition is the compositional mean of the sample;
we detail the notations for this scenario in the following.
At the compositional mean ($\pmb{x}_{0}$),
the within-person subcomposition, 
$\pmb{x}{^{(w)}_{0}}$, becomes the neutral element of the simplex,
$\pmb{1}_D = \pazocal{C}(1, 1, \ldots, 1) = (\kappa/D, \kappa/D, \ldots, \kappa/D)$,
as there is no within-person variance at the compositional mean.
We denote the compositional mean and its corresponding $ilr$ coordinates as

\begin{equation}
\begin{aligned}
\pmb{x}_{0}
&=
\pmb{x}^{(b)}_{0} \oplus \pmb{1}_{D}
=\pmb{x}^{(b)}_{0} \\
\pmb{z}_{0}
&=
\pmb{z}_{0}^{(b)} + \pmb{0}
\hspace{0.3cm}=\pmb{z}_{0}^{(b)} \\
\end{aligned}
\end{equation}
where 
$$
\pmb{x}^{(b)}_{0} =
  \pazocal{C}
  \left(
  {x}^{(b)}_{10},
  \ldots,
  {x}^{(b)}_{d0},
  \ldots,
  {x}^{(b)}_{d'0},
  \ldots,
  {x}^{(b)}_{D0}
  \right)
$$
and
$$
\pmb{z}_{0}^{(b)} = \text{ilr}(\pmb{x}^{(b)}_{0})
$$
We refer to the predicted outcome by the complete compositional predictor at the compositional mean as $\hat{y}_{0}$, expressed as

\begin{equation}
\begin{aligned}
  \hat{y}_{0} 
  &=
  \hat{\beta}_{0j} +
  \sum_{k = 1}^{D-1} \hat{\beta}_k               z{^{(b)}_{k0}} +
  \sum_{k = 1}^{D-1} \hat{\beta}_{(k + D - 1)}   z{^{(w)}_{k0}}
  \\
  &=
  \hat{\beta}_{0j} +
  \sum_{k = 1}^{D-1} \hat{\beta}_k               z{^{(b)}_{k0}} + 0 
  \\
  &=
  \hat{\beta}_{0j} +
  \sum_{k = 1}^{D-1} \hat{\beta}_k               z{^{(b)}_{k0}} \\
\end{aligned}
\end{equation}
We can perform the substitution analysis at different level of variability (e.g., between- and within-person) in the multilevel composition.

\subsubsection{Between-person Substitution Analysis}
We denote the two compositional parts involved in a given between-person pairwise substitution as 
$x{^{(b)}_{d0}}$ and $x{^{(b)}_{d'0}}$. Here, $d$ refers to the compositional part of the reference composition (e.g., compositional mean, $\pmb{x}^{(b)}_{0}$) that is reallocated a fixed unit $t$ from, and $d'$ refers to the part of the reference composition that is reallocated the same fixed unit $t$ to.
The between-person reallocation of a fixed unit $t$ from $x{^{(b)}_{d0}}$ to $x{^{(b)}_{d'0}}$
(i.e., adding $t$ to $x{^{(b)}_{d'0}}$ and subtracting $t$ from $x{^{(b)}_{d0}}$ simultaneously)
around the compositional mean $\pmb{x}_{0}$ is
\begin{equation}
  \begin{aligned}
    x{^{{(b)}'}_{d}}  &= x{^{(b)}_{d0}}  - t \\
    x{^{{(b)}'}_{d'}} &= x{^{(b)}_{d'0}} + t
  \end{aligned}
\end{equation}
where $d' \neq d \in \{1, \ldots, D\}$,  $t$ is the reallocated change 
(e.g., minutes/1440 if $\kappa=1440$), and $0 < t < \min \left\{ x^{(b)}_{d}, \kappa - x^{(b)}_{d'} \right\}$. 
Keeping the remaining compositional parts constant, the new $D$-part composition 
$\pmb{x}{^{{(b)}'}_{(d-d')}}$ can be expressed as
\begin{equation}
  \begin{aligned}
    \pmb{x}{^{{(b)}'}_{(d-d')}}
      & =
    \pazocal{C}(
    x{^{(b)}_{10}},
    \ldots,
    x{^{{(b)}'}_{d}},
    \ldots,
    x{^{{(b)}'}_{d'}},
    \ldots,
    x{^{(b)}_{D0}}) \\
      & =
    \pazocal{C}(
    x{^{(b)}_{10}},
    \ldots,
    (x{^{(b)}_{d0}}  - t),
    \ldots,
    (x{^{(b)}_{d'0}} + t),
    \ldots,
    x{^{(b)}_{D0}})
  \end{aligned}
\end{equation}
where ${(d-d')}$ denotes the reallocation of unit $t$ from the $d$ to the $d'$ compositional part relative to the reference composition.
The predicted outcome at the between-person reallocation is

\begin{equation}
  \begin{aligned}
    \hat{y}^{{(b)}'}_{(d-d')}
    =
    \hat{\beta}_{0j} +
    \sum_{k = 1}^{D-1} \hat{\beta}_k               z{^{{(b)}'}_{k0}}\\
  \end{aligned}
\end{equation}
where $z{^{{(b)}'}_{k0}}$ indicates the new between-person $ilr$ coordinates
resulted from the between-person reallocation in the composition (i.e., $\pmb{x}{^{{(b)}'}_{(d-d')}}$) and
$z{^{(w)}_{k0}}$ (within-person $ilr$ coordinates) remains unchanged.
The predicted difference in the outcome, $\Delta{\hat{y}^{(b)}}_{(d-d')}$, 
for the reallocation between the compositional mean and the reallocated composition at between-person level is therefore
\begin{equation}
    \Delta{\hat{y}^{(b)}}_{(d-d')} = \hat{y}^{{(b)}'}_{(d-d')} - \hat{y}_{0}.
\end{equation}
For models where the link function is the identity function, this becomes:

\begin{equation}
  \begin{aligned}
    \Delta{\hat{y}^{(b)}}_{(d-d')} & = \hat{y}^{{(b)}'}_{(d-d')} - \hat{y}_{0} \\
    & =
    \left(\hat{\beta}_{0j} +
    \sum_{k = 1}^{D-1} \hat{\beta}_k z{^{{(b)}'}_{k0}} \right)
    -
    \left(\hat{\beta}_{0j} +
    \sum_{k = 1}^{D-1} \hat{\beta}_k z{^{{(b)}}_{k0}} \right) \\
    & = \sum_{k = 1}^{D-1} \hat{\beta}_k \left(z{^{{(b)}'}_{k0}} - {z^{{(b)}}_{k0}} \right)
  \end{aligned}
\end{equation}

\subsubsection{Within-person Substitution Analysis}
The reallocation of a fixed amount $t$ between two 
compositional parts at the within-person level (from $x{^{(w)}_{d0}}$ to $x{^{(w)}_{d'0}}$)
around the compositional mean $\pmb{x}_{0}$ is
\begin{equation}
  \begin{aligned}
    x{^{{(w)}'}_{d}}  = x{^{(w)}_{d0}} - t  = 1  - t \\
    x{^{{(w)}'}_{d'}} = x{^{(w)}_{d'0}} + t = 1 + t
  \end{aligned}
\end{equation}
The new composition showing the within-person level reallocation of $t$ is
\begin{equation}
  \begin{aligned}
    \pmb{x}{^{{(w)}'}_{(d-d')}} 
      & =
    \pazocal{C}(
    x{^{(b)}_{10}},
    \ldots,
    x{^{(b)}_{d0}}          \pmb{\cdot}  x{^{{(w)}'}_{d}},
    \ldots,
    x{^{(b)}_{d'0}}         \pmb{\cdot}  x{^{{(w)}'}_{d'}},
    \ldots,
    x{^{(b)}_{D0}}) \\
      & =
    \pazocal{C}(
    x{^{(b)}_{10}},
    \ldots,
    x{^{(b)}_{d0}}          \pmb{\cdot}  (1 - t),
    \ldots,
    x{^{(b)}_{d'0}}         \pmb{\cdot}  (1 + t),
    \ldots,
    x{^{(b)}_{D0}}) \\
  \end{aligned}
\end{equation}
The predicted outcome for the within-person reallocation becomes
\begin{equation}
  \begin{aligned}
    \hat{y}^{(w)'}_{(d-d')}
    =
    \hat{\beta}_{0j} +
    \sum_{k = 1}^{D-1} \hat{\beta}_k               z{^{(b)}_{k0}} +
    \sum_{k = 1}^{D-1} \hat{\beta}_{(k + D - 1),j} z{^{{(w)}'}_{k0}}
  \end{aligned}
\end{equation}
where the $z{^{(b)}_{k0}}$ remains the same as the reference between-person $ilr$ coordinates, 
whereas the $z{^{{(w)}'}_{k0}}$ is the new within-person $ilr$ coordinates,
denoting the change in within-person $ilr$ coordinates relative to the compositional mean.
Thus, the predicted difference in the outcome associated with a reallocation across the compositional parts around the compositional mean at the within-person level, $\Delta{\hat{y}^{(w)}_{(d-d')}}$, is

\begin{equation}
\begin{aligned}
        \Delta{\hat{y}^{(w)}_{(d-d')}} 
    &= \hat{y}^{{(w)}'}_{(d-d')} - \hat{y}_{0} \\
    &=
    \left(
    \hat{\beta}_{0j} +
    \sum_{k = 1}^{D-1} \hat{\beta}_k               z{^{(b)}_{k0}} +
    \sum_{k = 1}^{D-1} \hat{\beta}_{(k + D - 1),j} z{^{{(w)}'}_{k0}}
    \right) \\
    &-
    \left(\hat{\beta}_{0j} +
    \sum_{k = 1}^{D-1} \hat{\beta}_k z{^{{(b)}}_{k0}} \right) \\
    &=
    \sum_{k = 1}^{D-1} \hat{\beta}_{(k + D - 1),j} z{^{{(w)}'}_{k0}}
\end{aligned}
\end{equation}

\begin{table*}[!htbp]
\caption{Steps to Perform Bayesian Multilevel Compositional Substitution Analysis.}
    \centering
    \small
    {\renewcommand{\arraystretch}{1.55}% for the vertical padding
    \begin{tabular*}{\textwidth}{@{\extracolsep{\fill}}l@{\extracolsep{\fill}}}
      \toprule
      Step and Notation  \\
      \midrule
\makecell[l]{1. Select a reference composition} \\ 
\makecell[c]{$\pmb{x}_{0}$}                     \\
                
\makecell[l]{2. Decompose into its between and within levels}   \\
\makecell[c]{$\pmb{x}{^{(b)}_{0}}$ and $\pmb{x}{^{(w)}_{0}}$} \\
                
\makecell[l]{3. Re-express composition as $ilr$ coordinates}      \\ 
\makecell[c]{$\pmb{z}{^{(b)}_{0}}$ and $\pmb{z}{^{(w)}_{0}}$} \\
                
\makecell[l]{4. Estimate the outcome by the complete composition at the reference composition} \\
\makecell[c]{$\hat{y}_{0} =
             \hat{\beta}_{0j} +
             \sum_{k = 1}^{D-1} \hat{\beta}_k               z{^{(b)}_{k0}} +
             \sum_{k = 1}^{D-1} \hat{\beta}_{(k + D - 1)}   z{^{(w)}_{k0}}$} \\

\textbf{A. Between substitution} \\
\makecell[l]{5A. Calculate the new composition for the reallocation at the \textit{between-person} level} \\
\makecell[c]{$\pmb{x}{^{{(b)}'}_{(d-d')}}$} \\

\makecell[l]{6A. Re-express the new composition as $ilr$ coordinates} \\
\makecell[c]{$\pmb{z}{^{{(b)}'}_{(d-d')}}$ and $\pmb{z}{^{{(w)}'}_{(d-d')}}$} \\

\makecell[l]{7A. Estimate the outcome at the \textit{between-person} reallocation} \\
\makecell[c]{$\hat{y}^{(b)'}_{(d-d')} = 
             \hat{\beta}_{0j} + 
             \sum_{k = 1}^{D-1} \hat{\beta}_k z{^{(b)'}_{k(d-d')}} + 
             \sum_{k = 1}^{D-1} \hat{\beta}_{(k + D - 1)} z{^{(w)}_{k(d-d')}}$} \\

\makecell[l]{8A. Estimate the difference in outcome between the \textit{between-person} reallocation and the reference} \\
\makecell[c]{$\Delta{\hat{y}^{(b)}_{(d-d')}} = \hat{y}^{(b)'}_{(d-d')} - \hat{y}_{0}$} \\ 

\textbf{B. Within substitution} \\
\makecell[l]{5B. Calculate the new composition for the reallocation at the \textit{within-person} level} \\
\makecell[c]{$\pmb{x}{^{{(w)}'}_{(d-d')}}$} \\

\makecell[l]{6B. Re-express the new composition as $ilr$ coordinates} \\
\makecell[c]{$\pmb{z}{^{{(b)}'}_{(d-d')}}$ and $\pmb{z}{^{{(w)}'}_{(d-d')}}$} \\

\makecell[l]{7B. Estimate the outcome for the \textit{within-person} reallocation} \\
\makecell[c]{$\hat{y}^{(w)'}_{(d-d')} = 
            \hat{\beta}_{0j} + 
            \sum_{k = 1}^{D-1} \hat{\beta}_k z{^{(b)}_{k(d-d')}} + 
            \sum_{k = 1}^{D-1} \hat{\beta}_{(k + D - 1)} z{^{{(w)}'}_{k(d-d')}}$} \\ 

\makecell[l]{8B. Estimate the difference in outcome between the \textit{within-person} reallocation and the reference} \\
\makecell[c]{$\Delta{\hat{y}^{(w)}_{(d-d')}} = \hat{y}^{(w)'}_{(d-d')} - \hat{y}_{0}$} \\

      \bottomrule
      \end{tabular*}} \\
\label{tab-submodel}
\end{table*}

\section{Software Implementation}
We implemented this method in a free, open-source, 
easy-to use \textbf{R} package \textit{multilevelcoda} \parencite{multilevelcoda, le2024b}.
The \textbf{R} package \textit{multilevelcoda} is built on \textit{brms} \parencite{brms, burkner2017} and 
\textbf{Stan} \parencite{stan2023}, which are easily accessible to lay users.
The focus of \textit{multilevelcoda} is on a streamlined and efficient workflow from 
dealing with raw multilevel compositional data, performing log-ratio transformations, 
estimating Bayesian multilevel models and the associated substitution analyses, 
and visualising final results. The \textbf{R} package supports generalised (non-)linear multivariate multilevel 
models using full Bayesian statistical inference. Models can treat compositions as predictors, 
outcomes, or both. Substitution analyses are currently supported for composition as a predictor with a univariate outcome. Substitution analyses for multivariate outcomes, including compositions as outcomes, are planned in the future. For details, see Table \ref{tab-features}.

\begin{table*}[!htbp]
  \caption{Supported Model Types and Substitution Analyses in \texttt{multilevelcoda}.}
  \centering\small
    {\renewcommand{\arraystretch}{1.25}% for the vertical padding
    \setlength\tabcolsep{0.25pt}
    \begin{tabular*}{\linewidth}{@{\extracolsep{\fill}} lccc}
\toprule
Bayesian model types & \makecell[c]{Compositional \\ predictor} & \makecell[c]{Compositional \\ outcome} & \makecell[c]{Substitution 
\\analysis} \\
\midrule
\makecell[l]{Single-level,\\univariate normal} & yes & - & yes \\
\makecell[l]{Single-level,\\multivariate normal} & yes & - & no\textsuperscript{\textdagger} \\
\makecell[l]{Single-level,\\univariate non-linear} & yes & - & yes \\
\makecell[l]{Single-level,\\multivariate non-linear} & yes & yes\textsuperscript{*} & no\textsuperscript{\textdagger}  \\
\makecell[l]{Multilevel,\\univariate normal} & yes & - & yes  \\
\makecell[l]{Multilevel,\\multivariate normal} & yes & - & no\textsuperscript{\textdagger} \\
\makecell[l]{Multilevel,\\univariate non-linear} & yes & - & yes \\
\makecell[l]{Multilevel,\\multivariate non-linear} & yes & yes\textsuperscript{*} & no\textsuperscript{\textdagger} \\
\bottomrule
  \end{tabular*}} \\
  \raggedright{\textit{Notes.} \textsuperscript{*}models with compositional outcomes can include compositional predictors. \textsuperscript{\textdagger}to be implemented.}
\label{tab-features}
\end{table*}

\begin{figure*}[!htbp]
  \caption{Workflow for Bayesian Multilevel Models with Compositional Predictors and Substitution Analysis using package \textbf{multilevelcoda}.}
  \centering
  \tikzstyle{process} = [rectangle, rounded corners, text width=6cm, minimum height=1.5cm, text centered, draw=black, thick]
  \tikzstyle{line} = [thick,-,>=stealth]
  \tikzstyle{arrow} = [thick,->,>=stealth]
  
  \begin{tikzpicture}[node distance = 3cm]
  \node (pro1) [process, align=center]                
  {\texttt{complr()}      \\ Compute multilevel compositional data and log-ratio transformations};
  \node (pro2) [process, align=center, below of=pro1] 
  {\texttt{brmcoda()}      \\ Fit Bayesian multilevel models for compositional data};
  \node (pro3) [process, align=center, below of=pro2] 
  {\texttt{substitution()} \\ Estimate Bayesian multilevel compositional substitution };

  \draw [arrow] (pro1) -- (pro2);
  \draw [arrow] (pro2) -- (pro3);

  \end{tikzpicture}
\label{fig-workflow}
\end{figure*}

\section{Real Data Study}
\label{rdatstudy}
We now demonstrate a real-data example application of this method in 
modelling compositional predictors, using the workflow outlined in Figure \ref{fig-workflow}.
The objectives of this study are to 1) examine the association between the 24h behaviours and sleepiness,
and 2) estimate the changes in sleepiness associated with the time reallocations across behaviours 
at both the between-person and within-person levels.

\subsection{Method}
\subsubsection{Data}
The data come from three studies with similar daily intensive designs and repeated measures:
Activity, Coping, Emotions, Stress, and Sleep (ACES, \textit{N} = 187);
Diet, Exercise, Stress, Emotions, Speech, and Sleep (DESTRESS, \textit{N} = 78); and
Stress and Health Study (SHS, \textit{N} = 96). 
Study materials are available on the Open Science framework for ACES (\url{https://doi.org/10.17605/OSF.IO/H5497}\nocite{aces}), 
DESTRESS (\url{https://doi.org/10.17605/OSF.IO/QM63W}\nocite{destress}), and 
SHS (\url{https://doi.org/10.17605/OSF.IO/TZ48Y}\nocite{shs}).
This data set has the structure found in typical applications of 
multilevel analysis in psychological research 
(i.e., daily observations nested within individuals).
For the purposes of this illustration, we used complete data of 345 individuals with repeated measurements of sleepiness and
24h time use separated into five behaviours: total sleep time, time awake in bed, 
MVPA, LPA, and SB.
Behaviours were recorded via an actigraph for 7-15 days and 
scored using the \textit{GGIR} \textbf{R} package \parencite{ggir1, ggir2, ggir3, ggir4, ggir5}.
Sleepiness was a single item and self-reported 3-4 times daily, 
which was averaged to obtain the daily level of sleepiness.
Study procedures have been described previously \parencite{le2022} and approved by the Monash University Human Research Ethics Committee (ACES \#8245, DESTRESS \#12637, SHS \#17281). 
Data are available from the corresponding authors upon request.

\subsubsection{Analytical Approach}
The 24h behaviours make up a 5-part composition ($D$ = 5), 
which corresponds to set of 4 ($D-1$) $ilr$ coordinates.
Days with missing data and zero values of any behaviours were excluded,
as missing data and zeros result in undefined $ilr$ coordinates.
The between- and within-person $ilr$ coordinates 
were constructed using the SBP shown in Table \ref{tab-sbp}, 
which formed the coordinates shown in Equation \ref{coords}.
The coordinates represent the relative information of the composition as follows
\begin{equation}
\begin{aligned}
z{^{(b)}_{1    \pmb{\cdot} j}}
&=
\ln
\begin{bmatrix}
\dfrac
{(\text{Sleep}^{(b)}_{\pmb{\cdot} j} \cdot \text{Awake in bed}^{(b)}_{\pmb{\cdot} j})^{\sqrt{\sfrac{3}{10}}}}
{(\text{MVPA}^{(b)}_{\pmb{\cdot} j} \cdot \text{LPA}^{(b)}_{\pmb{\cdot} j} \cdot \text{SB}^{(b)}_{\pmb{\cdot} j})^{\sqrt{\sfrac{2}{15}}}} 
\end{bmatrix}
\\
z{^{(b)}_{2    \pmb{\cdot} j}}
&=
\ln
\begin{bmatrix}
\dfrac
{(\text{Sleep}^{(b)}_{\pmb{\cdot} j})^{\sqrt{\sfrac{1}{2}}}}
{(\text{Awake in bed}^{(b)}_{\pmb{\cdot} j})^{\sqrt{\sfrac{1}{2}}}} 
\end{bmatrix}
\\
z{^{(b)}_{3    \pmb{\cdot} j}}
&=
\ln
\begin{bmatrix}
\dfrac
{(\text{MVPA}^{(b)}_{\pmb{\cdot} j})^{\sqrt{\sfrac{2}{3}}}}
{(\text{LPA}^{(b)}_{\pmb{\cdot} j} \cdot \text{SB}^{(b)}_{\pmb{\cdot} j})^{\sqrt{\sfrac{1}{6}}}}
\end{bmatrix}
\\
z{^{(b)}_{4    \pmb{\cdot} j}}
&=
\ln
\begin{bmatrix}
\dfrac
{(\text{LPA}^{(b)}_{\pmb{\cdot} j})^{\sqrt{\sfrac{1}{2}}}}
{(\text{SB}^{(b)}_{\pmb{\cdot} j})^{\sqrt{\sfrac{1}{2}}}} 
\end{bmatrix}
\\
\end{aligned}
\end{equation}
and
\begin{equation}
\begin{aligned}
z{^{(w)}_{1ij}}
&=
\ln
\begin{bmatrix}
\dfrac
{(\text{Sleep}^{(w)}_{ij} \cdot \text{Awake in bed}^{(w)}_{ij})^{\sqrt{\sfrac{3}{10}}}}
{(\text{MVPA}^{(w)}_{ij} \cdot \text{LPA}^{(w)}_{ij} \cdot \text{SB}^{(w)}_{ij})^{\sqrt{\sfrac{2}{15}}}} 
\end{bmatrix}
\\
z{^{(w)}_{2ij}}
&=
\ln
\begin{bmatrix}
\dfrac
{(\text{Sleep}^{(w)}_{ij})^{\sqrt{\sfrac{1}{2}}}}
{(\text{Awake in bed}^{(w)}_{ij})^{\sqrt{\sfrac{1}{2}}}} 
\end{bmatrix}
\\
z{^{(w)}_{3ij}}
&=
\ln
\begin{bmatrix}
\dfrac
{(\text{MVPA}^{(w)}_{ij})^{\sqrt{\sfrac{2}{3}}}}
{(\text{LPA}^{(w)}_{ij} \cdot \text{SB}^{(w)}_{ij})^{\sqrt{\sfrac{1}{6}}}}
\end{bmatrix}
\\
z{^{(w)}_{4ij}}
&=
\ln
\begin{bmatrix}
\dfrac
{(\text{LPA}^{(w)}_{ij})^{\sqrt{\sfrac{1}{2}}}}
{(\text{SB}^{(w)}_{ij})^{\sqrt{\sfrac{1}{2}}}}
\end{bmatrix}
\\
\end{aligned}
\end{equation}
where the between-person subcomposition is 
$\pmb{x}{^{(b)}_{ \pmb{\cdot} j}}    
= 
\pazocal{C} 
(
\text{Sleep}^{(b)}_{\pmb{\cdot} j}, 
\text{Awake in bed}^{(b)}_{\pmb{\cdot} j}, 
\text{MVPA}^{(b)}_{\pmb{\cdot} j}, 
\text{LPA}^{(b)}_{\pmb{\cdot} j}, 
\text{SB}^{(b)}_{\pmb{\cdot} j}
)
$
and the within-person subcomposition is
$
\pmb{x}{^{(w)}_{ij}}
= 
\pazocal{C} 
(
\text{Sleep}^{(w)}_{ij}, 
\text{Awake in bed}^{(w)}_{ij}, 
\text{MVPA}^{(w)}_{ij}, 
\text{LPA}^{(w)}_{ij}, 
\text{SB}^{(w)}_{ij}
)
$.
The $ilr$ coordinates represent the relative effects of behaviours 
(increasing in parts placed the numerator while decreasing in parts placed the denominator, by the same proportion), 
accounting for the constrained nature between behaviours within the 24h day. 
Specifically, across the between- and within-person levels, they represent the effects of
(1) increasing total time in sleep and awake in bed while proportionally decreasing total time in MVPA, LPA, and SB,
(2) increasing total sleep time while proportionally decreasing time awake in bed,
(3) increasing MVPA while proportionally decreasing LPA and SB, and
(4) increasing LPA while proportionally decreasing SB.

We considered a Bayesian multilevel model (denoted in Equation \ref{eq-mlm}).
The predictors were a total of 8 (4 between- plus 4 within-person) $ilr$ coordinates, representing the 5-part behaviour composition, and the outcome is \textit{next-day} sleepiness.
A group-level intercept by participants was included to account for non-independence.
The model was fitted with default, weakly informative priors, 4 chains, and 4 cores,
with 3000 iterations with the first 500 iterations treated as warmups
(total of 10000 post-warmup draws), using CmdStanR \parencite{cmdstanr} as back-end.
Model convergence was defined as all $\hat{R}$ < 1.05 and effective sample size (ESS) > 400 \parencite{vehtari2021}.
 
The default priors (Table \ref{tab-brmcoda-prior}) 
were designed to be weakly informative and play a minimal role in the computation of the posterior distribution, while maximising the influence of the data.
For the population-level effects, student's t distribution was used for 
the fixed intercept, and flat priors (improper priors over the reals) were used
for the parameters of the predictors.
Group-level effects also have their standard deviation parameters 
(i.e., random intercept and residual),
which were specified using
student's t distribution.
The priors for the standard deviation parameters are restricted to be non-negative and 
have a half student-t prior with 3 degrees of freedom and 
a scale parameter that depends on the standard deviation of the outcome. 
These priors are quickly overwhelmed by the data, 
but provide some regularisation to improve convergence and sampling efficiency. 
Prior sensitivity analysis was performed using importance sampling to estimate the properties of perturbed posteriors that result from power-scaling \parencite{kallioinen2024}. Following previous work \parencite{kallioinen2024}, the priors (except for the prior on the group-level intercept) were power-scaled by $\alpha$ = 0.5 (weaken), 1 (base), 2 (strengthen), and the extent to which the perturbed posteriors differ from the base posterior were evaluated both numerically (using Cumulative Jensen-Shannon distances) and visually (using Kernel density plot of power-scaled posterior draws). 

\begin{table*}[!htbp]
  \caption{Priors for Bayesian Multilevel Models with Compositional Predictors.}
  \centering\small
    {\renewcommand{\arraystretch}{1.25}% for the vertical padding
    \setlength\tabcolsep{0pt}
    \begin{tabular*}{\linewidth}{@{\extracolsep{\fill}} llll }
    \toprule
    & Parameter           & Prior  & Sensitivity\\
    \midrule
     \multicolumn{2}{l}{\textbf{Population-level (Fixed)}} && \\
    Intercept                               
    & $\gamma_0$                                         & student\_t(3, 1.7, 2.5) & 0.00      \\
     1\textsuperscript{st} between $ilr$  
    & $\beta{z{^{(b)}_{1    \pmb{\cdot} j}}}$            & flat                    & 0.00     \\
     2\textsuperscript{nd} between $ilr$  
    & $\beta{z{^{(b)}_{2    \pmb{\cdot} j}}}$            & flat                    & 0.00    \\
     3\textsuperscript{rd} between $ilr$  
    & $\beta{z{^{(b)}_{3    \pmb{\cdot} j}}}$            & flat                    & 0.00    \\
     4\textsuperscript{th} between $ilr$  
    & $\beta{z{^{(b)}_{4    \pmb{\cdot} j}}}$            & flat                    & 0.00     \\
     1\textsuperscript{st} within $ilr$   
    & $\beta{z{^{(w)}_{1ij}}}$                           & flat                    & 0.00    \\
     2\textsuperscript{nd} within $ilr$   
    & $\beta{z{^{(w)}_{2ij}}}$                           & flat                    & 0.00     \\
     3\textsuperscript{rd} within $ilr$   
    & $\beta{z{^{(w)}_{3ij}}}$                           & flat                    & 0.00     \\
     4\textsuperscript{th} within $ilr$   
    & $\beta{z{^{(w)}_{4ij}}}$                           & flat                    & 0.00     \\
    \midrule
    \multicolumn{2}{l}{\textbf{Group-level (Random)}} && \\
    Intercept                       
    & $\sigma_{u}$                                       & student\_t(3, 0, 2.5)   & 0.00\\
    Residual                        
    & $\sigma_{\varepsilon}$                             & student\_t(3, 0, 2.5)   & 0.00\\
    \bottomrule
  \end{tabular*}}  \\
  \raggedright{\textit{Notes.} Higher sensitivity values indicate greater sensitivity. Prior sensitivity above 0.05 indicates informative prior.}
\label{tab-brmcoda-prior}
\end{table*}

The Bayesian multilevel compositional substitution analysis (estimation procedure outlined in Table \ref{tab-submodel}) was then conducted for both between- and within-person levels, 
estimating the differences in \textit{next-day} sleepiness associated with the 
pairwise reallocation from 1 to 30 minutes between 24h behaviours.

Significance of individual parameters was assessed using the Bayesian 95\% posterior credible interval (CI), 
with 95\% CIs not containing 0 providing evidence that less than
5\% of the posterior distribution lies at 0 or on the opposite sign 
from the estimate.
All analyses were performed in R \parencite{R},
using packages \textit{multilevelcoda} v1.1.0 (model estimation,
workflow outlined in Figure \ref{fig-workflow}), 
\textit{brms} \parencite{brms, burkner2017} (back-ends for model estimation), 
\textit{priorsense}  \parencite[prior sensitivity,][]{kallioinen2024},
\textit{future} \parencite[parallel processing,][]{future},
and \textit{ggplot2} \parencite[results visualisation,][]{ggplot}.
Analysis code is available at: \url{https://github.com/florale/multilevelcoda-sim}.

\subsection{Results}
\subsubsection{Model Diagnostics}
\begin{figure*}[!htbp]
  \caption{Posterior densities depending on amount of prior power-scaling. 
  Overlapping lines indicates lower sensitivity, whereas wider gaps between lines indicate higher sensitivity.
  Estimates with high Pareto k (dashed lined) might be inaccurate.}
  \centering
  \includegraphics[width=\textwidth]{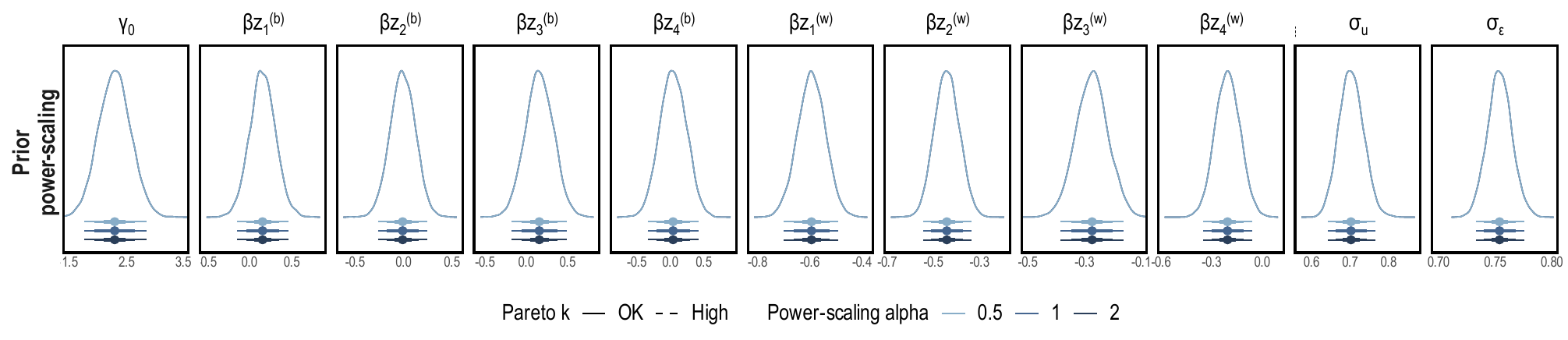} \\
\label{fig-priorsense}
\end{figure*}

The Bayesian multilevel model successfully converged (all $\hat{R} \approx $ 1.00 and ESS $\ge$ 2430).
Power-scaling posterior quantities (Figure \ref{fig-priorsense}) and sensitivity diagnostics 
(Table \ref{tab-brmcoda-prior}) indicated negligible prior sensitivity, 
indicating the minimal influence of the default, noninformative priors on the posterior.

\subsubsection{Bayesian Multilevel Model with Compositional Predictor}
Results from the Bayesian multilevel model predicting next-day sleepiness 
from a 24h behaviour composition 
are in Table \ref{tab-realstudy}, supporting the effects of 
all within-person $ilr$ coordinates
(indicated by 95\% CIs not containing 0s),
but not any between-person $ilr$ coordinates
(indicated by 95\% CIs containing 0s).
This demonstrated the associations between behaviours and next-day sleepiness
occurred only at the within-person level, but not the between-person level.
Overall, a unit increase in the $1^{\text{st}}$ within-person $ilr$ coordinate
(longer time spent on sleep behaviours than usual [total time in sleep and time awake in bed],
relative to wake behaviours [MVPA, LPA, and SB])
predicted -0.59 [95\% CI -0.70, -0.49] lower next-day sleepiness.
A unit increase in the $2^{\text{nd}}$ within-person $ilr$ coordinate 
(longer time sleeping than usual, 
relative to spending time staying awake in bed),
also predicted lower -0.44 [95\% CI -0.55, -0.34] next-day sleepiness.
Similarly,
a unit higher the $3^{\text{rd}}$ within-person $ilr$ coordinate 
(higher-than-usual MVPA, relative to LPA and SB)
and the $4^{\text{th}}$ within $ilr$ coordinate 
(higher-than-usual LPA relative to SB),
predicted lower sleepiness
(-0.27 [95\% CI -0.39, -0.16] and -0.20 [95\% CI -0.35, -0.06], respectively).

\begin{table*}[ht]
\caption{Bayesian Multilevel Model with Compositional Predictor Examining the Associations of the 24-hour Sleep-Wake Behaviours and Sleepiness.}
  \centering
  \small
  {\renewcommand{\arraystretch}{1.25}% for the vertical padding
  \begin{tabular*}{\textwidth}{@{\extracolsep{\fill}}llc@{\extracolsep{\fill}}}
    \toprule
    Parameter & Interpretation & \makecell{Posterior mean \\ {[95\% credible intervals]}}\\
    \midrule
    \textbf{Between-person level} & &\\
    $\beta{z{^{(b)}_{1    \pmb{\cdot} j}}}$  & 
    \makecell[l]{Longer sleep and awake in bed, \\relative to MVPA, LPA, and SB on average}
                   & $\begin{matrix} 0.16 \\ [-0.15, 0.46] \end{matrix}$ \\
                   
    $\beta{z{^{(b)}_{2    \pmb{\cdot} j}}}$  & 
    \makecell[l]{Longer sleep, \\ relative to awake in bed on average}
                   & $\begin{matrix} -0.01 \\ [-0.27, 0.25] \end{matrix}$ \\
                   
    $\beta{z{^{(b)}_{3    \pmb{\cdot} j}}}$  &
    \makecell[l]{Longer MVPA, \\ relative to LPA and SB on average}
                   & $\begin{matrix} 0.16 \\ [-0.17, 0.49] \end{matrix}$ \\
                   
    $\beta{z{^{(b)}_{4    \pmb{\cdot} j}}}$  &
    \makecell[l]{Longer LPA, \\ relative to SB on average}
                   & $\begin{matrix} 0.04 \\ [-0.36, 0.43] \end{matrix}$ \\
     \midrule
     \textbf{Within-person level} & &\\
     $\beta{z{^{(w)}_{1ij}}}$ & \makecell[l]{ Longer-than-usual sleep and awake in bed, \\relative to MVPA, LPA, and SB on a given day}
                   & $\begin{matrix} -0.59^\ast \\ [-0.69, -0.49] \end{matrix}$ \\
     $\beta{z{^{(w)}_{2ij}}}$ & \makecell[l]{Longer-than-usual sleep, \\relative to awake in bed on a given day}
                   & $\begin{matrix} -0.44^\ast \\ [-0.55, -0.34] \end{matrix}$ \\
     $\beta{z{^{(w)}_{3ij}}}$ & \makecell[l]{Longer-than-usual MVPA, \\ relative to LPA and SB on a given day}
                   & $\begin{matrix} -0.27^\ast \\ [-0.39, -0.16] \end{matrix}$ \\
     $\beta{z{^{(w)}_{4ij}}}$ & \makecell[l]{Longer-than-usual LPA, \\ relative to SB within level on a given day}
                   & $\begin{matrix} -0.20^\ast \\ [-0.35, -0.06] \end{matrix}$ \\

     \bottomrule
    \end{tabular*}} \\
  \raggedright{\textit{Notes.} MVPA = moderate-to-vigorous physical activity, LPA = light physical activity, SB = sedentary behaviour. $^\ast$95\% credible intervals not containing 0.}
\label{tab-realstudy}
\end{table*}

\subsubsection{Bayesian Multilevel Compositional Substitution Analysis}
\begin{figure*}[!htbp]
  \centering
  \caption{Estimated Differences in Sleepiness for 1-30 Minute Reallocations of 24-hour Behaviours. MVPA = moderate-to-vigorous physical activity, LPA = light physical activity, SB = sedentary behaviour. The panels represent the pairwise reallocations. For example, the top left panel shows reallocation between LPA and total sleep time, where positive values on the x-axis (e.g., +30 minute) indicate reallocations of from LPA to total sleep time, whereas negative values (e.g., -30) indicate reallocations of from total sleep time to LPA.}
  \includegraphics[width=\textwidth]{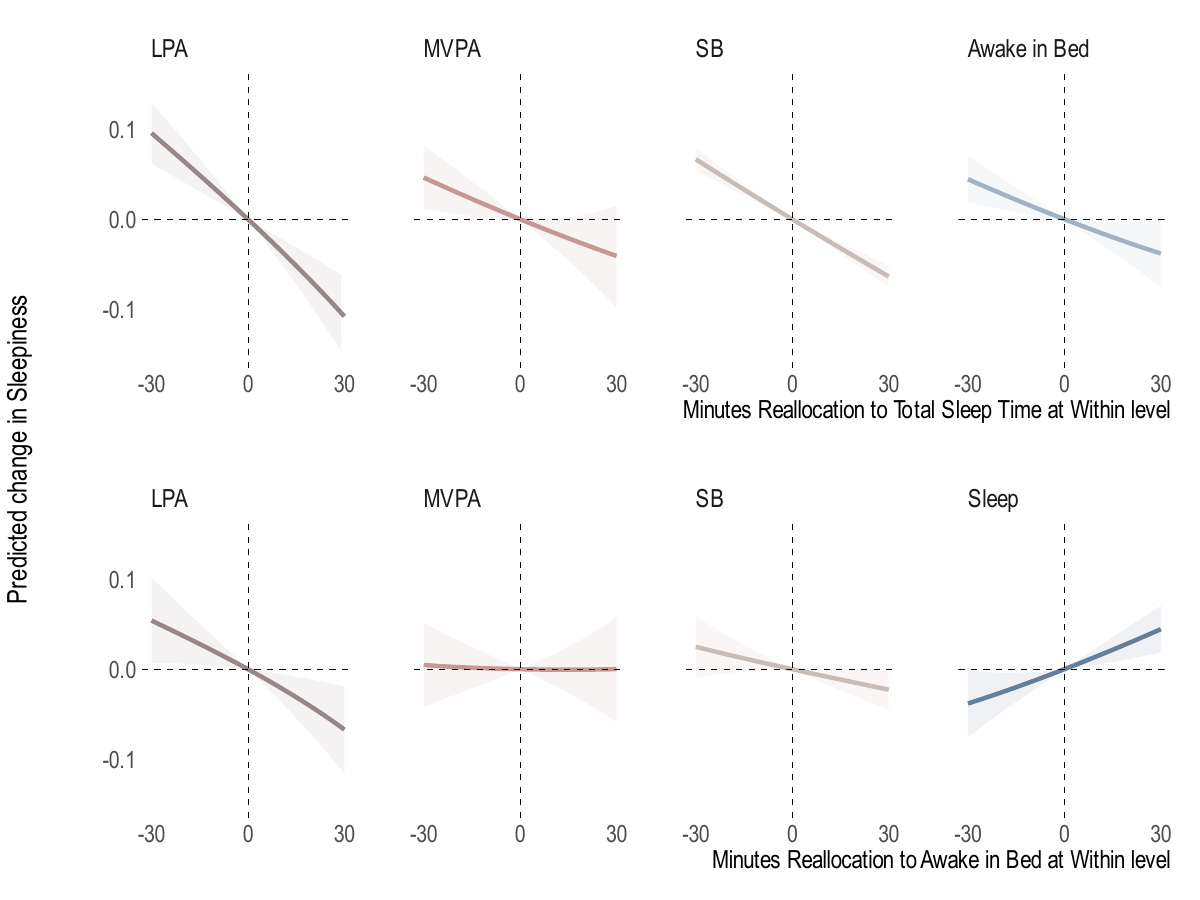}
\label{fig-realdatsubplot}
\end{figure*}

Bayesian multilevel compositional substitution analysis showed that
reallocation of time between 24h behaviours predicted changes in sleepiness
at the within-person level, but not the between-person level.
Individuals who slept longer-than-usual at the expense of any behaviours, except MVPA, 
at within level, experienced lower levels of next-day sleepiness.
However, when individuals sacrificed their sleep on a given day for any other behaviours (i.e., including MVPA), they experienced a higher level of sleepiness.
Additionally, individuals who spent longer time in LPA
at the expense of time awake in bed on a given day, 
also experienced a higher level of sleepiness the next day, and vice versa.
Results of the substitution analysis for 30-minute reallocations are
in Table \ref{tab-rdsubmodel}.
For brevity, we present only the statistically significant results for the reallocation from 1 to 30 minutes of total sleep time and awake in bed, respectively, 
in Figure \ref{fig-realdatsubplot}.

\begin{table*}[!htbp]
\caption{Bayesian Multilevel Compositional Substitution Analysis Estimating the Difference in Sleepiness Associated with Reallocation of 30 minutes across 24-hour Sleep-Wake Behaviours.}
 \centering
 \footnotesize
  {\renewcommand{\arraystretch}{1.25}% for the vertical padding
  \begin{tabular*}{\textwidth}{@{\extracolsep{\fill}}lccccc@{\extracolsep{\fill}}}
  \toprule
                & $\downarrow \text{Sleep}$
                & $\downarrow \text{Awake in bed}$
                & $\downarrow \text{MVPA}$
                & $\downarrow \text{LPA}$
                & $\downarrow \text{SB}$ \\
      \midrule
      \textbf{Between-person level} &&&&& \\
      $\uparrow \text{Sleep}$
                & -
                & $\begin{matrix} -0.05 \\ [-0.15, 0.05] \end{matrix}$
                & $\begin{matrix} -0.05	\\ [-0.22, 0.12] \end{matrix}$
                & $\begin{matrix}  0.04	\\ [-0.08, 0.16] \end{matrix}$
                & $\begin{matrix}  0.01 \\ [-0.02, 0.04] \end{matrix}$                 \\

      $\uparrow \text{Awake in bed}$
                & $\begin{matrix}  0.03	\\ [-0.04, 0.09] \end{matrix}$
                & -
                & $\begin{matrix} -0.02 \\ [-0.20, 0.17] \end{matrix}$
                & $\begin{matrix}  0.07 \\ [-0.06, 0.20] \end{matrix}$
                & $\begin{matrix}  0.04 \\ [-0.02, 0.10] \end{matrix}$                 \\

      $\uparrow \text{MVPA}$
                & $\begin{matrix}  0.02 \\ [-0.08, 0.13] \end{matrix}$
                & $\begin{matrix} -0.03 \\ [-0.17, 0.11] \end{matrix}$
                & -
                & $\begin{matrix}  0.07	\\ [-0.14, 0.27] \end{matrix}$
                & $\begin{matrix}  0.04 \\ [-0.06, 0.14] \end{matrix}$                 \\

      $\uparrow \text{LPA}$
                & $\begin{matrix} -0.03 \\ [-0.13, 0.06] \end{matrix}$
                & $\begin{matrix} -0.08 \\ [-0.21, 0.05] \end{matrix}$
                & $\begin{matrix} -0.08 \\ [-0.32, 0.17] \end{matrix}$
                & -
                & $\begin{matrix} -0.02 \\ [-0.11, 0.07] \end{matrix}$                 \\

      $\uparrow \text{SB}$
                & $\begin{matrix} -0.01 \\ [-0.04, 0.02] \end{matrix}$
                & $\begin{matrix} -0.06 \\ [-0.16, 0.03] \end{matrix}$
                & $\begin{matrix} -0.06 \\ [-0.22, 0.11] \end{matrix}$
                & $\begin{matrix}  0.03 \\ [-0.09, 0.15] \end{matrix}$
                & -                                            \\
      \midrule
      \textbf{Within-person level} &&&&& \\
      $\uparrow \text{Sleep}$
                & -
                & $\begin{matrix} -0.04      \\ [-0.08,  0.00] \end{matrix}$
                & $\begin{matrix} -0.04      \\ [-0.10,  0.01] \end{matrix}$
                & $\begin{matrix} -0.11^\ast \\ [-0.15, -0.06] \end{matrix}$
                & $\begin{matrix} -0.06 ^\ast\\ [-0.08, -0.05] \end{matrix}$           \\

      $\uparrow \text{Awake in bed}$
                & $\begin{matrix}  0.04^\ast \\ [ 0.02,  0.7]  \end{matrix}$
                & -
                & $\begin{matrix}  0.00      \\ [-0.06,  0.06]  \end{matrix}$
                & $\begin{matrix} -0.07^\ast \\ [-0.12, -0.02]  \end{matrix}$
                & $\begin{matrix} -0.02      \\ [-0.04,  0.00]  \end{matrix}$          \\

      $\uparrow \text{MVPA}$
                & $\begin{matrix}  0.05^\ast \\ [ 0.01,  0.08]  \end{matrix}$
                & $\begin{matrix}  0.00      \\ [-0.04,  0.05]  \end{matrix}$
                & -
                & $\begin{matrix} -0.06      \\ [-0.14,  0.01]  \end{matrix}$
                & $\begin{matrix} -0.02      \\ [-0.05,  0.01]  \end{matrix}$          \\

      $\uparrow \text{LPA}$
                & $\begin{matrix}  0.10^\ast \\ [ 0.06,  0.13]  \end{matrix}$
                & $\begin{matrix}  0.05^\ast \\ [ 0.01,  0.10]  \end{matrix}$
                & $\begin{matrix}  0.05      \\ [-0.03,  0.13]  \end{matrix}$
                & -
                & $\begin{matrix}  0.03      \\ [-0.01,  0.06]  \end{matrix}$           \\

      $\uparrow \text{SB}$
                & $\begin{matrix}  0.07^\ast \\ [ 0.05,  0.08]  \end{matrix}$
                & $\begin{matrix}  0.03      \\ [-0.01,  0.06]  \end{matrix}$
                & $\begin{matrix}  0.02      \\ [-0.03,  0.08]  \end{matrix}$
                & $\begin{matrix} -0.04      \\ [-0.09,  0.00]  \end{matrix}$
                & -                                            \\
      \bottomrule
    \end{tabular*}}
    \raggedright{\textit{Notes.} MVPA = moderate-to-vigorous physical activity, LPA = light physical activity, SB = sedentary behaviour. Values are posterior means and 95\% credible intervals. $^\ast$95\% credible intervals not containing 0.}
    \label{tab-rdsubmodel}
\end{table*}

\section{Simulation Study}
\label{simstudy}
In a series of simulation studies,
we investigated the performance of the Bayesian multilevel model with compositional predictor and Bayesian multilevel compositional substitution analysis in parameter recovery. 
Our simulation study was based on the real data study, where the objective was 
to examine the association between 24h behaviour composition and sleepiness.

\subsection{Method}
\subsubsection{Simulation Conditions}
We created a range of simulation conditions including different values for
the number of clusters ($J$),
cluster size ($I$),
the number of compositional parts ($D$),
and the magnitude of sample variability
(assessed by the group-level intercept variance $\sigma^{2}_{u}$ and residual variance $\sigma^{2}_{\varepsilon}$).
The values for the number of clusters and cluster sizes were informed 
by a systematic review and meta-analyses on daily sleep and physical activity \parencite{atoui2021}.
Given the different number of compositional parts used in existing studies,
we constructed models with different numbers of possible compositional parts
and assessed their performances using different sets of ground truth values.
Finally, we examined the influences of sample variability,
including group-level intercept variance ($\sigma^{2}_{u}$) and
residual variance ($\sigma^{2}_{\varepsilon}$)
on the estimation of our models. 
Table \ref{tab-simcond} summarises the factors and their levels considered
in this simulation study.
The combination of these factors resulted in a total of 240 scenarios.
For each cell of the simulation design, 2000 replications were generated ($n_{sim}=2000$), resulting in 
$4 [I] \times 4 [J] \times 3 [D] \times 5 [\sigma] \times 2000 =$ 480 000
data sets to be analysed. 

\begin{table*}[!htbp]
  \caption{Factors and Their Levels for the Simulation Study.}
    \centering\small
    {\renewcommand{\arraystretch}{1.25}% for the vertical padding
    \setlength\tabcolsep{2.5pt}
    \begin{tabular*}{\linewidth}{@{\extracolsep{\fill}} lll}
      \toprule
     Factor                                             & Notation   & Levels \\ 
      \midrule
        Number of clusters                              & J          & 3, 5, 7, 14         \\ 
        Cluster size                                    & I          & 30, 50, 360, 1200   \\ 
        \makecell[l]{Number of \\compositional parts}   & D          & 3, 4, 5             \\ 
        \multirow{5}{*}{\makecell[l]{Variance \\ (group-level intercept \\ and residual \\ variance)}}  
        & \multirow{5}{*}{\makecell{$\sigma^2_{u}$ and $\sigma^2_{\varepsilon}$}}
        & $\sigma^2_{u} = 1$ and $\sigma^2_{\varepsilon} = 1$,     \\
        && $\sigma^2_{u} = 1.5$ and $\sigma^2_{\varepsilon} = 0.5$, \\
        && $\sigma^2_{u} = 0.5$ and $\sigma^2_{\varepsilon} = 1.5$, \\
        && $\sigma^2_{u} = 1$ and $\sigma^2_{\varepsilon} = 0.5$,   \\
        && $\sigma^2_{u} = 1$ and $\sigma^2_{\varepsilon} = 1.5$   \\
      \bottomrule
      \end{tabular*}} \\
    \raggedright{\textit{Notes.} $\sigma^2_{u}$ = group-level intercept variance $\sigma^2_{\varepsilon}$ = residual variance.}
    \label{tab-simcond}
\end{table*}

\subsubsection{Data Generation} 
The simulation procedure to generate data sets resembling the data structure used in real data study was as follows. The group-level intercept $u_{0j}$ was generated from $\text{Normal}(0, \sigma^{2}_{u})$.
The design matrices of the predictors, the between-person $ilr$ ($\pmb{z}{^{(b)}_{\pmb{\cdot} j}}$) and within-person $ilr$ ($\pmb{z}{^{(w)}_{ij}}$)
corresponding to 5-part composition of 24h behaviours (total sleep time, time in bed awake, MVPA, LPA, and SB) were generated, respectively, as follows:
$$\pmb{z}{^{(b)}_{\pmb{\cdot} j}}  \sim \text{MVNormal}(\pmb{\mu}^{\pmb{z}{^{(b)}_{\pmb{\cdot} j}}}, \pmb{\Sigma}^{\pmb{z}{^{(b)}_{\pmb{\cdot} j}}})$$
and 
$$\pmb{z}{^{(w)}_{ij}}  \sim \text{MVNormal}(\pmb{\mu}^{\pmb{z}{^{(w)}_{ij}}}, \pmb{\Sigma}^{\pmb{z}{^{(w)}_{ij}}})$$
with values of the means and covariances informed by the data set used in the real data study. Compositional data were then generated by inverse-transforming the 4-dimension $ilr$ coordinates.
At this step, when necessary,
the 4-part and 3-part compositions were created by collapsing variables.
The 4-part composition was obtained by collapsing total sleep time and wake in bed to a single variable named sleep.
The 3-part composition was created by collapsing MVPA and LPA
to a single variable named physical activity.
These compositions were transformed again to $ilr$ coordinates for model estimation. The outcome vector $\pmb{y}$ was generated from a normal distribution following Equation \ref{eq-mlm}:
$$
  \pmb{y} \sim \text{Normal}(
    \beta_{0j} +
    \sum_{k = 1}^{D-1}\beta_k            z^{(b)}_{k \pmb{\cdot} j} +
    \sum_{k = 1}^{D-1}\beta_{(k + D - 1)}z^{(w)}_{kij},
    \hspace{0.2cm}
    \sigma^{2}_{\varepsilon})
$$
with the values for the population-level parameters set to be close
to those found in the real data study.

\subsubsection{Parameters}
The primary parameters of interest in the simulation study are
the parameters of the Bayesian multilevel models, including the
population-level parameters: the intercept ($\gamma_0$), 
the between-person and within-person $ilr$ coordinates ($\beta$s),
and the group-level parameters:
group-level intercept ($\sigma_{u}$) and 
residual error ($\sigma_{\varepsilon}$).
For the Bayesian multilevel compositional substitution analysis,
estimation of predicted differences in outcome at between-level ($\Delta{\hat{y}^{(b)}_{(d-d')}}$) and 
within-level ($\Delta{\hat{y}^{(w)}_{(d-d')}}$)
were evaluated for all possible pairwise substitution between compositional parts, 
totalling to $2 \times D \times (D-1)$ parameters.

\subsubsection{Evaluation Criteria}
Model performance of 2000 replications across 240 conditions was evaluated using the following criteria:

\begin{itemize}[leftmargin=*, labelsep=2mm]
\item  Quality of the MCMC-based sampling procedure of the Bayesian multilevel model were 
       assessed using the proportion of replications that sufficiently converged \parencite[$\hat{R} < 1.05$, ][]{vehtari2021}
       and had no divergent transitions. 
       ESS was investigated 
       both at the bulk of the distribution (e.g., for the mean or median) and 
       in the tails (e.g., for posterior interval estimates and inferences about extreme quatiles).
       Any parameters with ESS $<$ 400 indicated sampling inefficiency and required further diagnostics \parencite{vehtari2021}.
\item  Quality of model performance was evaluated in terms of accuracy in parameter estimates and inference, 
       using three performance measures: bias, coverage, and bias-eliminated coverage \parencite{morris2019}. 
       Monte Carlo standard errors were used to calculate 95\% confidence intervals.
\end{itemize}

\subsubsection{Analytical Approach}
Using package \textit{multilevelcoda}, each simulated data set
was analysed using a Bayesian multilevel model with a group-level intercept
to predict \textit{next-day} sleepiness from 
the $D$-part behaviour composition, expressed as a total of $2(D-1)$ 
between- and within-person $ilr$ coordinates. The same model settings (e.g., priors, iterations, cores, chains) as the real data study were used.
The Bayesian multilevel compositional substitution analysis was then conducted to estimate the difference in sleepiness for 30-minute reallocation.
The simulation study results were summarised using package \textit{rsimsum} \parencite{rsimsum}
and visualised using package \textit{ggplot2} \parencite{ggplot}.
Reproducible material for this study is available at: 
\url{https://github.com/florale/multilevelcoda-sim}.

\subsection{Results}
We found minimal effects of 
certain simulation conditions on model estimation.
Therefore, for brevity,
the descriptive statistics of the simulation results
of the Bayesian multilevel compositional models and its associated substitution analyses were collapsed across
240 conditions and summarised in Table \ref{tab-simdes}.

\begin{table*}[!htbp]
\caption{Descriptive Statistics of Results from the Simulation Study.}
  \centering\small
  {\renewcommand{\arraystretch}{2}% for the vertical padding
  \setlength\tabcolsep{0pt}
  \begin{tabular*}{\linewidth}{@{\extracolsep{\fill}} lll }
    \toprule
                & \makecell[l]{Bayesian \\ Compositional \\ Multilevel \\ Models}
                & \makecell[l]{Bayesian \\ Compositional \\ Substitution \\ Analyses} \\
     \midrule
     \makecell[l]{Number of \\ divergent transitions}      & \makecell[l]{0.01 \\ (0, 134)}             & -  \\
      $\hat{R}$                                            & \makecell[l]{1.00 \\ (1.00, 1.07)}         & -  \\
      Bulk-ESS                                             & \makecell[l]{6193.83 \\ (52.06, 27047.59)} & -  \\
      Tail-ESS                                             & \makecell[l]{5600.04 \\ (107.91, 9465.94)} & -  \\
      \midrule
      Bias                                                 & \makecell[l]{0.00 \\ (-0.09, 0.05)} 
                                                           & \makecell[l]{0.00 \\ (-0.03, 0.04)} \\

      Coverage                                             & \makecell[l]{0.95 \\ (0.93, 0.97)}
                                                           & \makecell[l]{0.95 \\ (0.93, 0.97)} \\ 
      \makecell[l]{Bias-eliminated \\  
      Coverage}                                            & \makecell[l]{0.95 \\ (0.93, 0.97)}
                                                           & \makecell[l]{0.95 \\ (0.93, 0.97)} \\
    \bottomrule
    \end{tabular*}} \\
  \raggedright{\textit{Notes.} ESS = effective sample size. Values are mean and range.}
\label{tab-simdes}
\end{table*}

\subsubsection{Quality of Estimation Procedure}
Divergences were observed in 1312 replications (0.27\%), of which predominantly have
small number of clusters (73.6\% $J$: 30), small cluster size (90.5\% $I$: 3), and large residual variation (97.6\% $\sigma^2_{\varepsilon}: 1.5$).
An additional 17 (0.00\%) replications had $\hat{R} > 1.05$, demonstrating convergence issues.
These replications were excluded for the evaluation of parameter estimates and inference.

In contrast, low bulk ESS was observed 
as sample size increased with large between-person heterogeneity and small within-person heterogeneity.
Particularly, 27 651 replications (5.76\%) had
bulk $\text{ESS} < 400$ for some parameters, of which predominantly have
large number of clusters (51.1\% $J$: 1200), large cluster size (70.1\% $I$: 14), and small residual variation (95.6\% $\sigma^2_{\varepsilon}: 0.5$).
The low ESS values under these conditions may be a technical difficulty 
posed by the MCMC sampling methods,
where small variation in the sample (i.e., $\sigma^2_{\varepsilon}$) cause
the sampler to produce higher within-chain correlation \parencite{betancourt2015}.
Additionally, the default non-centered parameterisation 
\parencite[i.e., separation of population parameters and individual parameters in the prior, ][]{papaspiliopoulos2007}
used in our model estimation procedure 
can be less efficient for large data sets and strong likelihood (i.e., small sample variability), 
compared to centered parameterisation \parencite{betancourt2015}.
Therefore, we conducted a case study 
(presented as a vignette in the \textbf{R} package \textit{multilevelcoda}) 
into a replication generated
using a 3-part composition, 1200 clusters and cluster size of 14, 
with large group-level intercept variation ($\sigma^2_{u} = 1.5$) and residual variation ($\sigma^2_{\varepsilon} = 0.5$).
This model produced low bulk ESS values for 4 out of 7 parameters.
Posterior predictive distributions were checked and 
two methods to improve the MCMC sampling were tested:
centered-parameterisation and increased iterations.
Results showed no evidence of non-convergence 
(e.g., poor mixing of chains or funnel degeneracy in the posterior).
Both reparameterisation or increasing iterations and warm-ups improved ESS,
with centered parameterisation showing
substantial gain of ESS for the same number of iterations.
A sensitivity analysis comparing the model performance with and without the replications with low ESS
revealed that ESS did not have an influence on
the quality of parameter estimates and inference.
Replications with low ESS were, therefore, kept in the subsequent evaluation 
of parameter estimates and inferences.

\subsubsection{Quality of Parameter Estimates and Inference}
Across the simulated conditions,
both Bayesian multilevel models and 
Bayesian multilevel compositional substitution analyses 
yielded negligible biases in the estimation of all parameters.
For Bayesian multilevel models with compositional predictors, bias had a mean of 0.00 and a range from -0.09 to 0.05.
For Bayesian multilevel compositional substitution analyses, bias had 
a mean of 0.00 and range from -0.03 to 0.04.
Both models had coverage and bias-eliminated coverage close to the nominal 95\% value,
with means of 0.95 and ranges from 0.93 to 0.97.

As the models performed consistently well across conditions,
for brevity, results for individual parameters estimated from a 5-part 
composition ($D$ = 5) and a medium level of modelled variance 
($\sigma^2_{u} = 1$ and $\sigma^2_{\varepsilon} = 1$)
under different conditions of the number of clusters ($J$) and cluster size ($I$) 
are reported in Appendix \ref{appendix}.
Full results are accessible via a dedicated shiny app 
we have included in our \textbf{R} package \textit{multilevelcoda}.

\section{Discussion}\label{discussion}
This paper presented a Bayesian approach to modelling multilevel compositional data,
with a focus on both within-person and between-person processes.
We described the theories underlying the data and models and 
illustrated how to perform this method in a real data application.
A simulation study
confirmed the overall good performance of both 
Bayesian multilevel models and multilevel compositional substitution analyses 
under different simulation scenarios. 
Multilevel compositional data are becoming increasingly common.
For example, EMAs and wearable devices to advance clinical and health science have blossomed in the last decade.
These methodologies, especially employed in intensive, longitudinal research, have enabled 
the full day of behaviours and experiences to be captured.
In the wake of such data abundance,
this innovative statistical method which appropriately address the data properties of 
multilevel composition can enhance psychological studies and lead to new insights.

Our empirical results demonstrated the usefulness of the proposed method in
examining how day-to-day 24h behaviours 
are associated with other daily experiences using EMA data.
We showed that the reallocation of time between 24h behaviours 
was associated with \textit{next-day} sleepiness, 
and that this association differed by behaviours involved in the reallocation 
(e.g., sleep at the expense of MVPA \textit{or} SB), 
and whether the effect occurred at the between-person or within-person level.
These findings
highlight the importance of addressing the multilevel and compositional nature of 24h behaviours, 
and any other data with such properties.

Results of the simulation study showed that the quality of estimation procedures was related to
sample size and variability.
Divergences were observed in a small number of models fitted with small sample sizes and large sample variability,
whereas inefficiency of MCMC sampling, indicated by the low ESS, was observed
in models fitted with large data sets and small sample variability.
The estimation procedure in the simulation study followed a common framework 
for MCMC sampling \parencite{betancourt2015, betancourt2017, brms},
and diagnosing and dealing with sampling inefficiency 
depends on the model of interest and specific applications.
Nevertheless, we suggest the following.
To eliminate divergences, we recommend
using data sets with more than 30 clusters with a cluster size of 3 ($N$ = 90).
Studies that have already collected data or have sampling limitations 
may consider adjusting the initial step size and target acceptance rate to 
assist the sampling departure and trajectories in model estimation,
such as setting the \enquote{adapt\_delta} control parameter 
to a higher value than the default when fitting the models \parencite{schad2021}.
Scenarios with convergence issues or sampling inefficiency, 
indicated by low ESS and high $\hat{R}$, 
may be improved by 
reparameterisation or increasing the number 
of warm-up iterations and/or the number of posterior draws.
We found that reparameterisation, in particular, yielded 
the most robust ESS for the same number of iterations.

Bayesian multilevel models and Bayesian multilevel compositional substitution analyses both 
successfully recovered all tested summary statistics, including 
population-level and group-level parameters, and residual error.
Unbiased estimates and excellent coverage 
were consistently observed
across all conditions of sample sizes,
compositional parts, and the magnitude of sample variability.
This performance was further not influenced by the efficiency of MCMC sampling.
For frequentist multilevel models, a minimum data with 30 clusters with a cluster size of 50 is recommended for models using likelihood estimation methods (either full maximum likelihood or restricted maximum likelihood) 
to achieve unbiased estimates \parencite{mcneish2016}.
Frequentist multilevel models with smaller sample sizes may require Kenward–Roger adjustment \parencite{kenward1997}.
In contrast,
we showed that multilevel models estimated using Bayesian MCMC sampling can achieve unbiased estimates 
for data with 30 clusters with a cluster size of 3,
and other studies have provided evidence for data with fewer than ten clusters \parencite{stegmueller2013, browne2006}.
Another important advantage of our method lies in the substitution analysis.
Using the posterior predictive distributions, the model can directly describe
the uncertainty of the estimated quantities (i.e., the predicted changes in outcomes),
eliminating the computational burden of relying on resampling techniques, 
such as bootstrapping.

As with other Bayesian methods, 
the estimation time required for the models presented in this study is considerable.
With more complex models, larger data sets, 
or when investigating model sensitivity, transforming parameterisation,
the amount of time and computational resources can become increasingly substantial.
However, we believe that the advantages associated with this method,
including accurate and unbiased parameter estimates, 
straightforward estimation procedure, and minimal convergence issues,
outweigh the time and computational cost.
Running models or chains in parallel on a computing cluster can help speed up model estimation process.
Our recommended software for working with multilevel compositional data, 
including \textit{multilevelcoda}, \textit{brms}, and \textbf{Stan}, all provide options for
using multiple CPU cores to run Bayesian models in parallel.

It is important to note that these models requires complete and non-zero data.
Zeros and missing data hamper the analysis of compositional data,
as the ilr transformation is essentially based on log-ratios.
Although dealing with zeros and missing data is outside the scope of this study,
previous studies have discussed the zero composition problem \parencite{smithson2024, martin2003},
and have provided a comparison of different strategies 
in dealing with zeros in compositional data  \parencite{rasmussen2020},
and multilevel missing data \parencite{ludtke2017}.
Log-ratio Expectation-Maximisation \parencite[][]{palarea2015} has been recommended for zero imputation 
as it preserves the relative structure (i.e., ratios) of composition \parencite{rasmussen2020}. 
Imputation strategy based on multivariate multilevel models \parencite{schafer2002, pan}
has been shown to produce valid inferences for 
multilevel models with missing data at the lowest level of the multilevel structure
\parencite{ludtke2017}, such as observations of 24h behaviours.

\subsection{Limitations and Future Directions}
The model presented in this study was a multilevel model with Gaussian distribution and 
a group-level intercept. We did not examine 
a maximal random-effect structure (i.e., both group-level intercept and group-level slopes),
but that is fully supported in \textit{multilevelcoda}. 
Future simulation studies can also evaluate the performance of the model for these other outcome
distributions frequently observed in psychological research,
such as Bernoulli (binary data, such as depression status), 
Poisson (count data, such as number of cigarettes smoked per day). Although \textit{multilevelcoda} allows other outcome
distributions in the Bayesian multilevel model, multilevel compositional substitution analysis for non-normal outcomes is not yet implemented and remains a future direction.

Three-level data structures (e.g., behaviours nested within people, who in turn are nested within hospitals) are less common than two-level data, but do occur in psychological research.
Similarly, data can be cross-classified with observations nested within non-hierarchical clusters.
Addressing more than two-level and cross-classified data structures is an 
important area for future research. One initial question to be answered is 
what is the best way to disaggregate effects with more than two-levels?
Research has only recently suggested an approach to disaggregating effects for 
cross-classified data \parencite{guo2024}, and to our knowledge, approaches to optimally 
disaggregate effects in more than two-level data structures has 
not yet been established. Currently, we suggest keeping the
aggregate multilevel composition for more than two-level data structures 
when modelling them using \textit{multilevelcoda}.

Findings provide support for the minimal influence of 
noninformative priors on the posteriors.
When default priors were used,
negligible prior sensitivity was found in the real data study, and posteriors successfully recovered the simulation population in location and interval coverage in the simulation study.
However, we did not include informative priors,
which generally becomes of greater importance the smaller the sample size.
Due to complexity of the models and the current limited knowledge about 24h behaviour composition 
and its association with other outcomes, setting informative priors is challenging at this point.
Prior elicitation requires sufficient quantitative theoretical and empirical knowledge on the topic,
which may be enabled in the future as applications of this method in substantive research increase,
allowing more evidence-based consensus about prior decisions.
In the current cases when the data reasonably inform the likelihood, 
researchers are recommended to use default priors and perform sensitivity analysis.
Further research is warranted to systematically investigate 
the influences of prior choices, 
particularly informative priors, 
on the posteriors for multilevel compositional models.

Finally, we focused on composition as a predictor. 
However, composition can be considered as an outcome. As we support such model in 
\textit{multilevelcoda}, future research may evaluate the 
performance of Bayesian multilevel models in this scenario.
Further, other fields of research, such as behavioural epidemiology, 
are increasingly interested in understanding within-person variability 
(e.g., changes of behavioural composition at follow-up 
relative to baseline predicting changes in health outcomes), 
yet methods are not well established.
Our method may be explored in such data sets to extend its impacts beyond psychology.
More tutorials detailing step-by-step analyses of example data sets in different areas
could help promote wider applications of this innovative method.

\subsection{Conclusion}
We introduced an elegant method that integrates three statistical frameworks:
compositional data analysis, multilevel modelling, and Bayesian inference.
The implementation of this method 
in an open-source \textbf{R} package, \textit{multilevelcoda}, with
a user-friendly setup that only requires the data, 
model formula and minimal specification of the analysis, 
speaks to the feasibility of modelling multilevel compositional data in a novel way.
As the availability of data with a multilevel compositional structure is growing,
we believe Bayesian multilevel compositional data analysis will be an increasingly important tool to advance psychological research.
We hope that our tutorial, evaluations through simulations, and recommendations, 
will motivate researchers to employ this method in their works and disciplines 
to obtain robust answers to scientific questions that otherwise would be inaccessible.

\pagebreak

%%%%%%%%%%%%%%
\printbibliography

\appendix
\section{Bias and Coverage of Individual Parameter Estimation from the Simulation Study} \label{appendix}

\begin{figure*}[!htbp]
\centering
  \caption{Bias of Bayesian Multilevel Models with Five-Part Compositional Predictor and Medium Level of Variance. 
  Parameters are population- and group-level parameters from Bayesian multilevel models. 
  Values are mean estimates and 95\% confidence intervals. 
  J = Number of clusters, I = Cluster size.}
  \includegraphics[height=0.8\textheight, keepaspectratio]{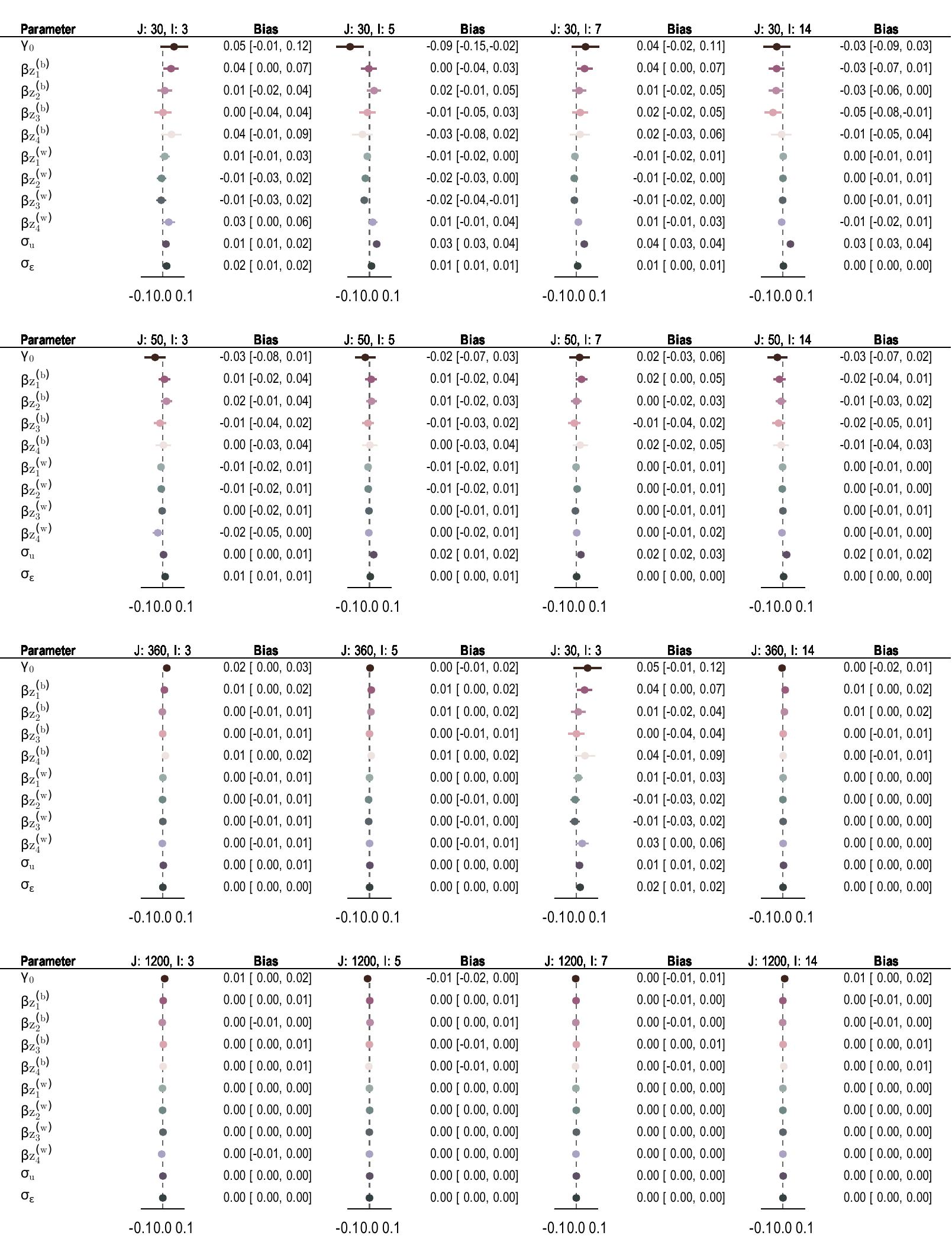} \\
\label{fig-brmcoda-bias}
\end{figure*}

\begin{figure*}[!htbp]
\centering
  \caption{Coverage of Bayesian Multilevel Models with Five-Part Compositional Predictor and Medium Level of Variance. 
  Parameters are population- and group-level parameters from Bayesian multilevel models.
  Values are mean estimates and 95\% confidence intervals.
  J = Number of clusters, I = Cluster size.}
  \includegraphics[height=0.8\textheight, keepaspectratio]{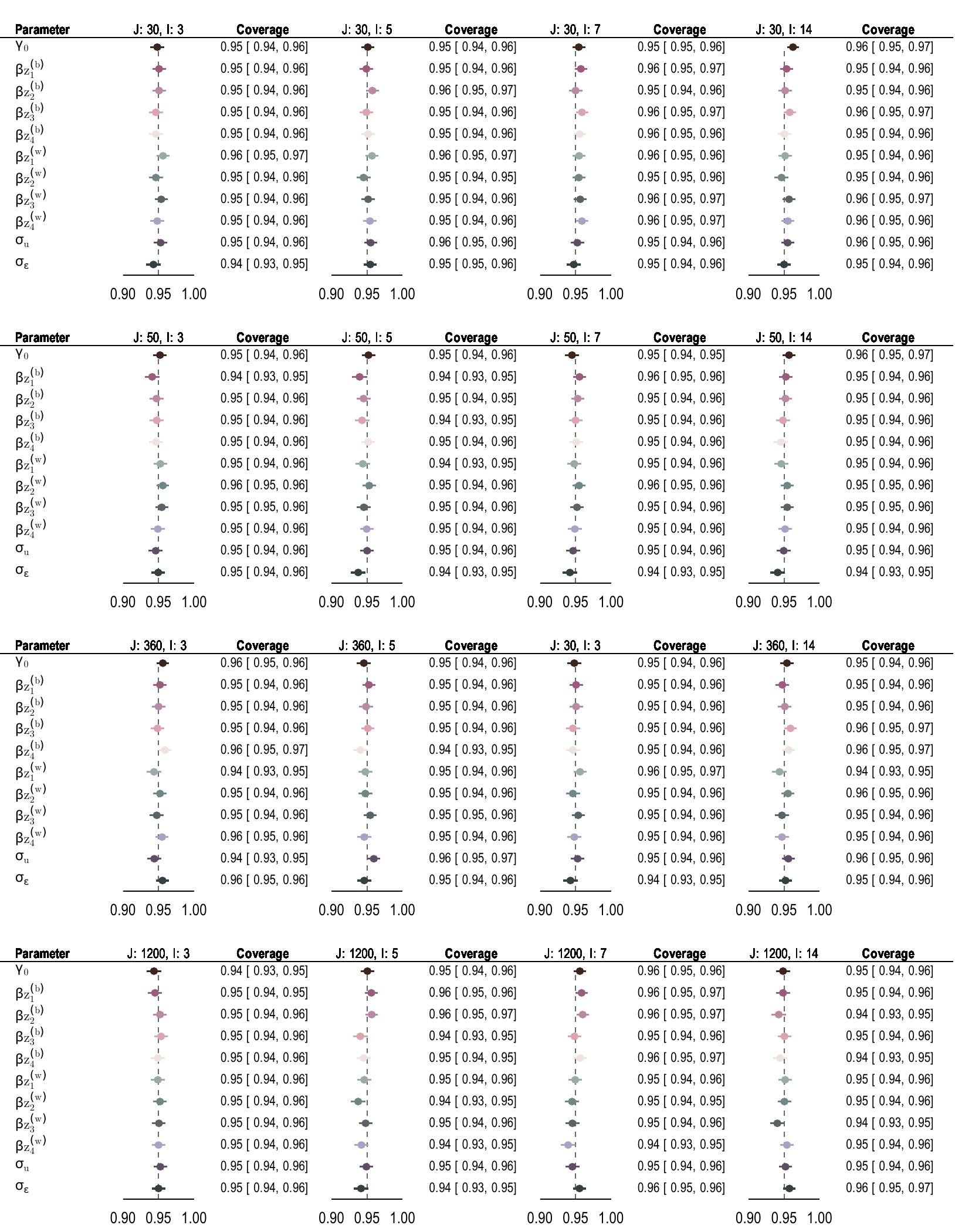} \\
\label{fig-brmcoda-cover}
\end{figure*}

\begin{figure*}[!htbp]
\centering
  \caption{Bias of Bayesian Multilevel Compositional Substitution Analysis with Five-Part Composition and Medium Level of Variance. 
  Parameters are predicted differences in outcome at between-level ($\Delta{\hat{y}^{(b)}_{(d-d')}}$) and within-level ($\Delta{\hat{y}^{(w)}_{(d-d')}}$), where ${(d-d')}$ denotes the reallocation of unit $t$ from the $d$ to the $d'$ compositional part relative to the compositional mean. For example, $(\text{MVPA}-\text{SB})$ means reallocation from MVPA to SB. 
  Values are mean estimates and 95\% confidence intervals. 
  MVPA = moderate-to-vigorous physical activity, LPA = light physical activity, SB = sedentary behaviour, TST = total sleep time, WAKE = Awake in bed. 
  J = Number of clusters, I = Cluster size.}
  \includegraphics[height=0.8\textheight, keepaspectratio]{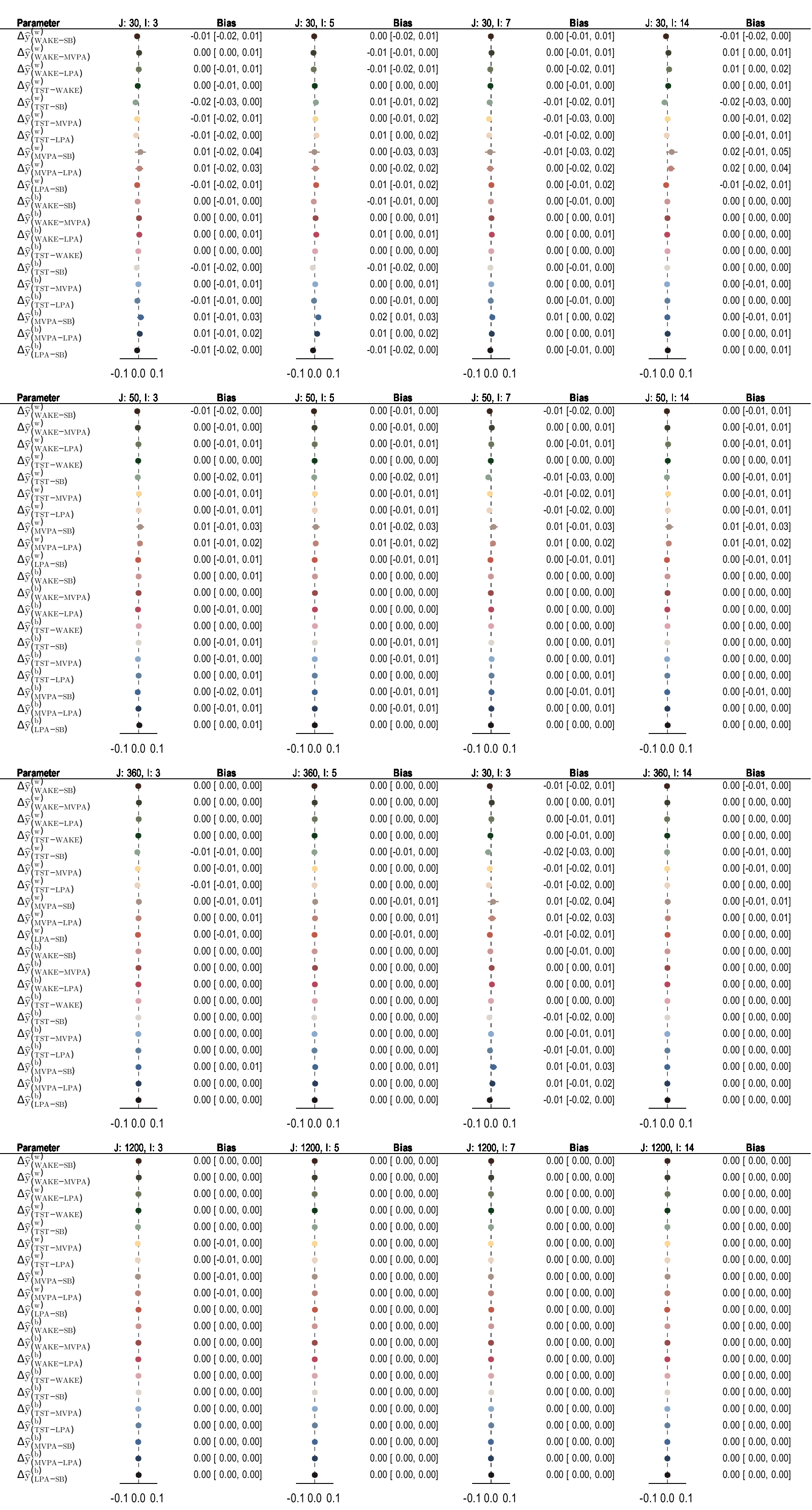} \\
\label{fig-sub-bias}
\end{figure*}

\begin{figure*}[!htbp]
\centering
  \caption{Coverage of Bayesian Multilevel Compositional Substitution Analysis with Five-Part Composition and Medium Level of Variance. 
  Parameters are predicted differences in outcome at between-level ($\Delta{\hat{y}^{(b)}_{(d-d')}}$) and within-level ($\Delta{\hat{y}^{(w)}_{(d-d')}}$), where ${(d-d')}$ denotes the reallocation of unit $t$ from the $d$ to the $d'$ compositional part relative to the compositional mean. For example, $(\text{MVPA}-\text{SB})$ means reallocation from MVPA to SB. 
  Values are mean estimates and 95\% confidence intervals. 
  MVPA = moderate-to-vigorous physical activity, LPA = light physical activity, SB = sedentary behaviour,  TST = total sleep time, WAKE = Awake in bed. 
  J = Number of clusters, I = Cluster size.}
  \includegraphics[height=0.8\textheight, keepaspectratio]{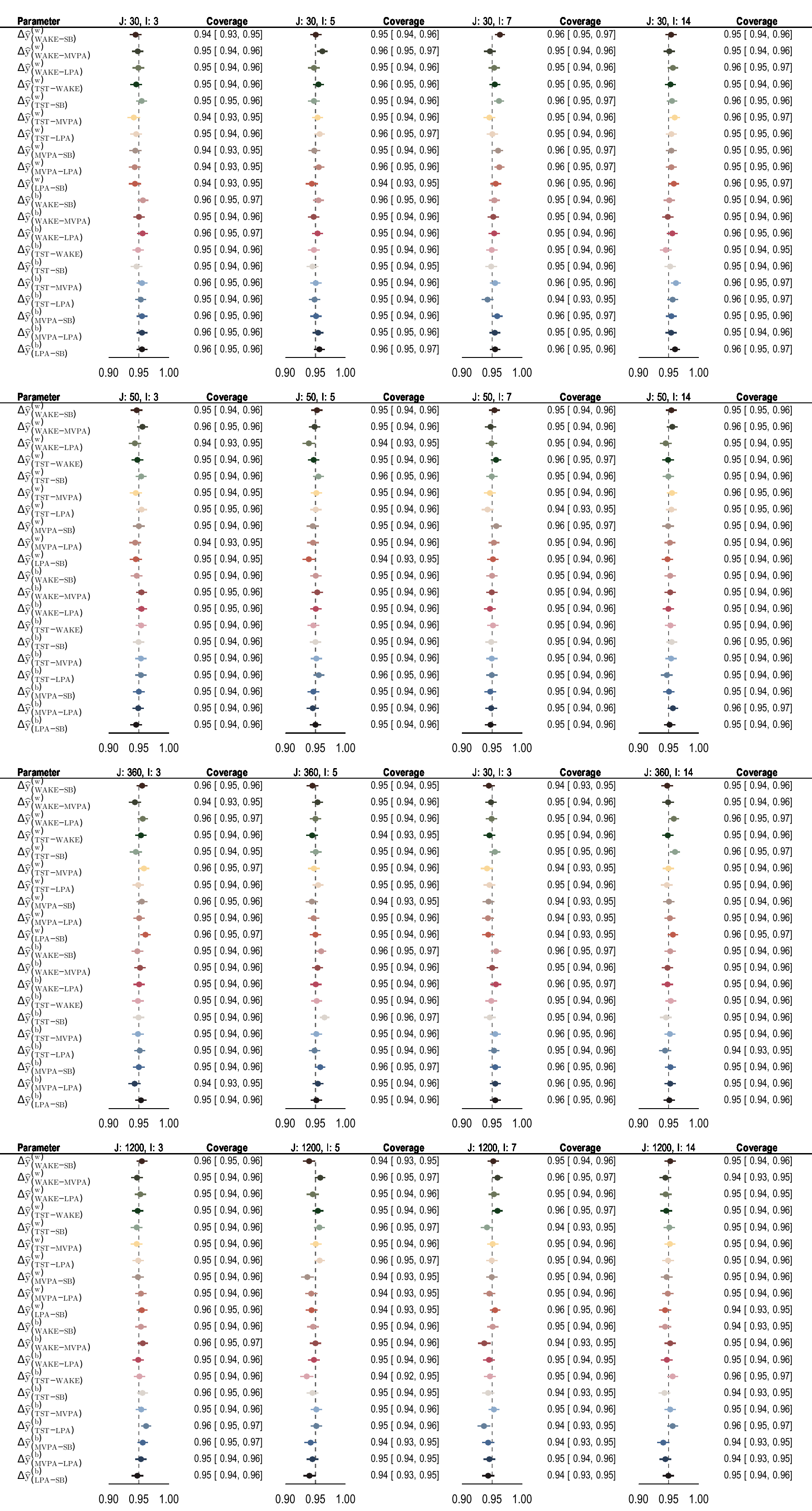} \\
\label{fig-sub-cover}
\end{figure*}

\end{document}